\documentclass{article}

\usepackage{amsmath,amsfonts,amssymb,amscd,latexsym,multicol,multirow}
\usepackage{caption,subcaption}
\usepackage{graphicx}
\usepackage{algorithm,algorithmic}
\usepackage{url}
\usepackage{appendix}
\usepackage[numbers]{natbib}

\usepackage[T1]{fontenc}
\usepackage[utf8]{inputenc}
\usepackage{authblk}

\newcommand{\mD}{\mathcal{D}}
\newcommand{\mM}{\mathcal{M}}
\newcommand{\vtheta}{\vec{\theta}}
\newcommand{\vpsi}{\vec{\psi}}
\newcommand{\vsigma}{\vec{\sigma}}

\title{Hierarchical Stochastic Model in Bayesian Inference: Theoretical Implications and Efficient Approximation\thanks{This work was supported by the European Research Council (Advanced Investigator Award no. 341117).}} 

\author[1]{S. Wu\thanks{Currently at The Institute of Statistical Mathematics, Japan.}}
\author[1]{P. Angelikopoulos}
\author[2]{J.~L. Beck}
\author[1]{P. Koumoutsakos}
\affil[1]{Computational Science and Engineering Laboratory, Claussiusstrasse 33, ETH-Zurich 8092, Switzerland.}
\affil[2]{Department of Mechanical and Civil Engineering, California Institute of Technology, Pasadena, CA 91102, USA.}

\begin{document}

\maketitle

\begin{abstract}
We classify two types of Hierarchical Bayesian Model found in the literature as Hierarchical Prior Model (HPM) and Hierarchical Stochastic Model (HSM). Then, we focus on studying the theoretical implications of the HSM. Using examples of polynomial functions, we show that the HSM is capable of separating different types of uncertainties in a system and quantifying uncertainty of reduced order models under the Bayesian model class selection framework. To tackle the huge computational cost for analyzing HSM, we propose an efficient approximation scheme based on Importance Sampling and Empirical Interpolation Method. We illustrate our method using two examples --- a Molecular Dynamics simulation for Krypton and a pharmacokinetic/pharmacodynamic model for cancer drug.\\
Key words: Hierarchical Bayesian, Importance Sampling, Empirical Interpolation Method, Molecular Dynamics, Pharmacokinetics
\end{abstract}

\section{Introduction}

Hierarchical Bayesian Model (HBM) is a powerful modeling tool extended from the classical Bayesian framework. It has been used to solve many difficult practical problems, such as modeling heterogeneous data, calibrating system with multiple objectives and inducing sparsity in a model. Its applications span across various fields, such as, social science, economics, physics, medical science, civil engineering, etc \cite{Woolrich:2004oq,FF+Perona:2005,Ballesteros:2014,Hahn:2015jk}. 

In a classical Bayesian setting, users define a stochastic model with parameters $\vtheta$ for a forward problem that predicts a quantity of interest and a prior distribution of $\vtheta$. When data is available, the Bayes' Theorem is used to solve the inverse problem by finding the posterior distribution of $\vtheta$. For a complex system in reality, we further parameterize the stochastic model and the prior with hyperparameters. This extra level of parameters provide extra flexibility to a model, but generally requires more data to reach a well-posed problem. HBM refers to model classes with such a hierarchy (multiple levels) of parameters. Here, we define two types of HBM commonly found in the literature: a Hierarchical Prior Model (HPM) that further parameterizes the prior, and a Hierarchical Stochastic Model (HSM) that further parameterizes the stochastic model (or known as the likelihood function when evaluated at a given data). 

HPM is a well-studied subject in the machine learning and compressive sensing community in the context of Sparse Bayesian Learning (SBL). Its applications include imaging \cite{Sato+et_al:2004, Calvetti:2008kq}, gene selection \cite{Bae:2004qy}, Bayesian compressive sensing \cite{Ji:2008fq, Huang+et_al:2014}, etc. It is an effective tool to solve ill-posed regression problems by adding extra constraints in the prior distribution under the Bayesian framework. The well-known Bayesian Optimization method can also be interpreted using this model structure. 
On the other hand, the origin of HSM can be traced back to the multilevel model in the statistics community. One classic application is to analyze test results collected from multiple schools \cite{Gelman+Hill:2006}. \cite{Tiao:1965if} and \cite{Hill:1965sf} are early publications on Bayesian analysis for multilevel regression, and Congdon \cite{Congdon:2010} summarizes recent developments on this topic. Since most of the computational demand in Bayesian analysis comes from evaluating the stochastic model, HSM has a significantly larger computational cost than HPM. Although recently, there is increasing amount of HSM applications, the diversity of HSM uses is not comparable to HPM. However, recent developments in parallel computing opens up new opportunities to apply HSM to more complicated problems. In this paper, we discuss important theoretical implications of HSM and demonstrate an efficient approximation scheme with practical applications.

Despite the frequent use of the terminology ``HBM" in the literature, the distinction between what we called the HSM and the HPM is often omitted. The overlapped usage of ``HBM" has even caused confusion. For example, Guha et al. \cite{Guha:2015qy} studied the HPM. Nevertheless, the authors used \cite{Gelman:2004bk} as a reference, which the type of HBM mentioned in \cite{Gelman:2004bk} is actually the HSM. Hence, in this paper, we begin with a comparison between HPM and HSM in Section \ref{sec:HBM}. Then, we turn our focus back to HSM and study its theoretical implications based on polynomial regression in Section \ref{sec:theory}. To tackle the high computational cost for analyzing HSM in practice, we propose an efficient approximation based on the idea of Importance Sampling and Empirical Interpolation Method \cite{Barrault:2004rz} in Section \ref{sec:approx}. Section \ref{sec:ex} includes two realistic examples using our approximation method. Finally, we give some concluding remarks in Section \ref{sec:con}.

% This particular usage of HSM is the most common one in the literature. For instance, Woolrich et al. \cite{Woolrich:2004oq} use the model for functional magnetic resonance imaging analysis and Ballesteros et al. \cite{Ballesteros:2014} use it for uncertainty quantification in structural dynamics. 

\section{Classification of HBM: HPM versus HSM}\label{sec:HBM}

For the ease of illustration, we begin with defining the notations that will be repeatedly used in this paper. To predict a quantity of interest $y$, one may have a stochastic forward model $F(x,\vtheta,\vec{\epsilon})$ with model parameters $\vtheta$ and the stochasticity part $\vec{\epsilon}$. This function defines the distribution $p(y|x,\vtheta)$. One of the most common example of $F(\cdot)$ is:
\begin{equation}\label{eq:basicSM}
	y = f(x,\vtheta) + \epsilon_y,
\end{equation}
where $\epsilon_y$ is chosen to follow a zero mean and $\sigma_y$ standard deviation Gaussian distribution, $N(0,\sigma_y)$, for computational convenience. Given a set of input and output data pairs $\mD = \{(\hat{x}_i,\hat{y}_i)|i,\dots,N_D\}$, the posterior distribution of $\vtheta$ is inferred by the Bayes' Theorem:
\begin{equation}\label{eq:BayesThm}
	p(\vtheta|\mD) = \frac{p(\mD|\vtheta)p(\vtheta)}{p(\mD)},
\end{equation}
where the likelihood $p(\mD|\vtheta) \triangleq p(\hat{y}_1,\dots,\hat{y}_{N_D}|\hat{x}_1,\dots,\hat{x}_{N_D},\vtheta)$ is the probability of observing the data values $\mD$ from the predictive model, the prior $p(\vtheta)$ is the initial belief of the values of $\vtheta$, and the evidence (or marginal likelihood) $p(\mD) \triangleq p(\hat{y}_1,\dots,\hat{y}_{N_D}|\hat{x}_1,\dots,\hat{x}_{N_D})$ is a critical term used in model selection \cite{Beck:2004qy,Beck:2010}. Then, a robust posterior prediction for an unobserved input-output pair $(x_0,y_0)$ can be obtained by:
\begin{equation}\label{eq:RobustPred}
	p(y_0|x_0,\mD) = \int \! p(y_0|x_0,\vtheta)p(\vtheta|\mD) \, \mathrm{d}\vtheta.
\end{equation}

\begin{figure}[h!]
	\centering
	\begin{subfigure}[t]{0.4\textwidth}
		\centering
		\includegraphics[width=\textwidth]{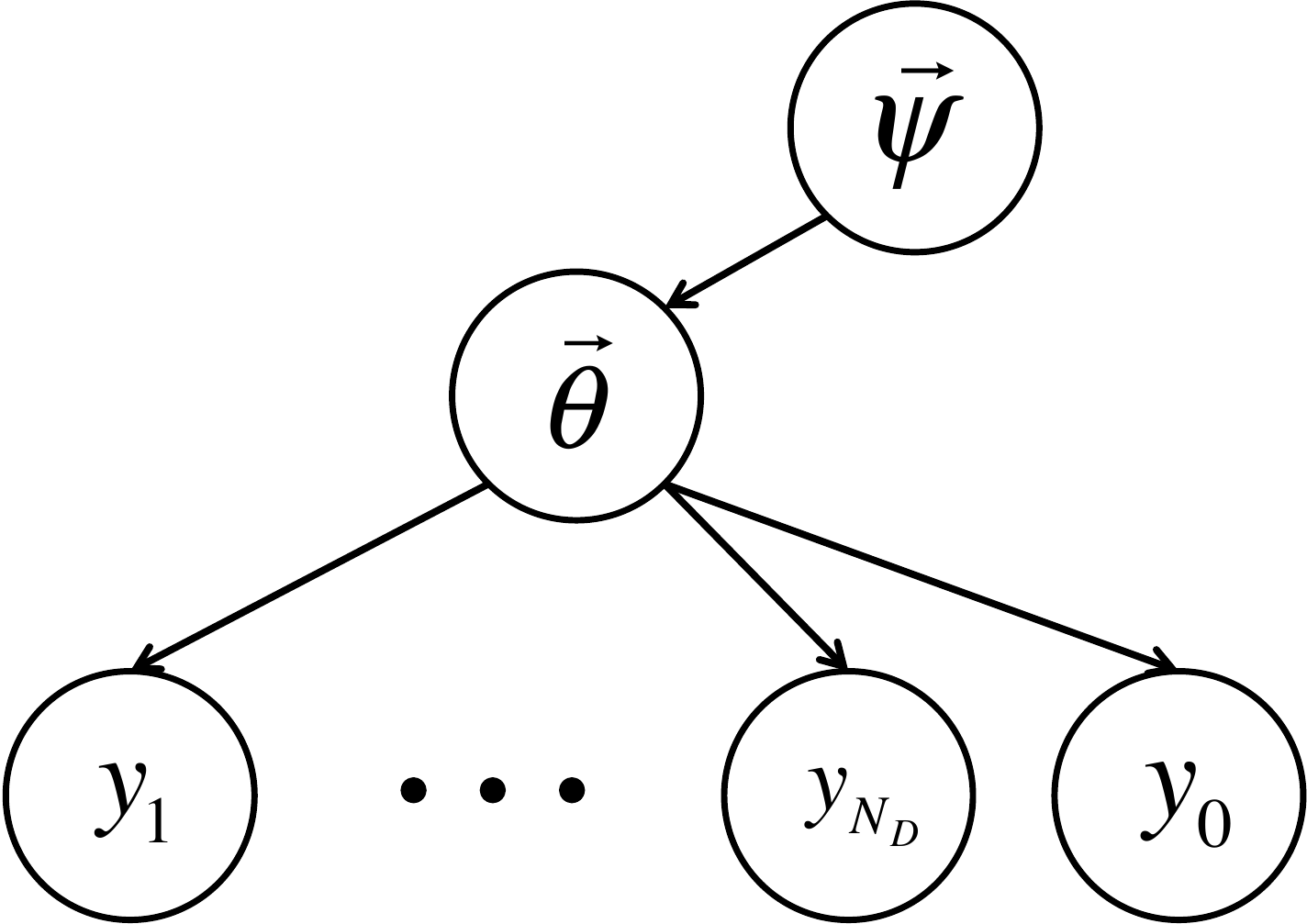}
		\caption{Hirerachical prior model}
		\label{fig:BN_NB}
	\end{subfigure}
	\qquad
	\begin{subfigure}[t]{0.4\textwidth}
		\centering
		\includegraphics[width=\textwidth]{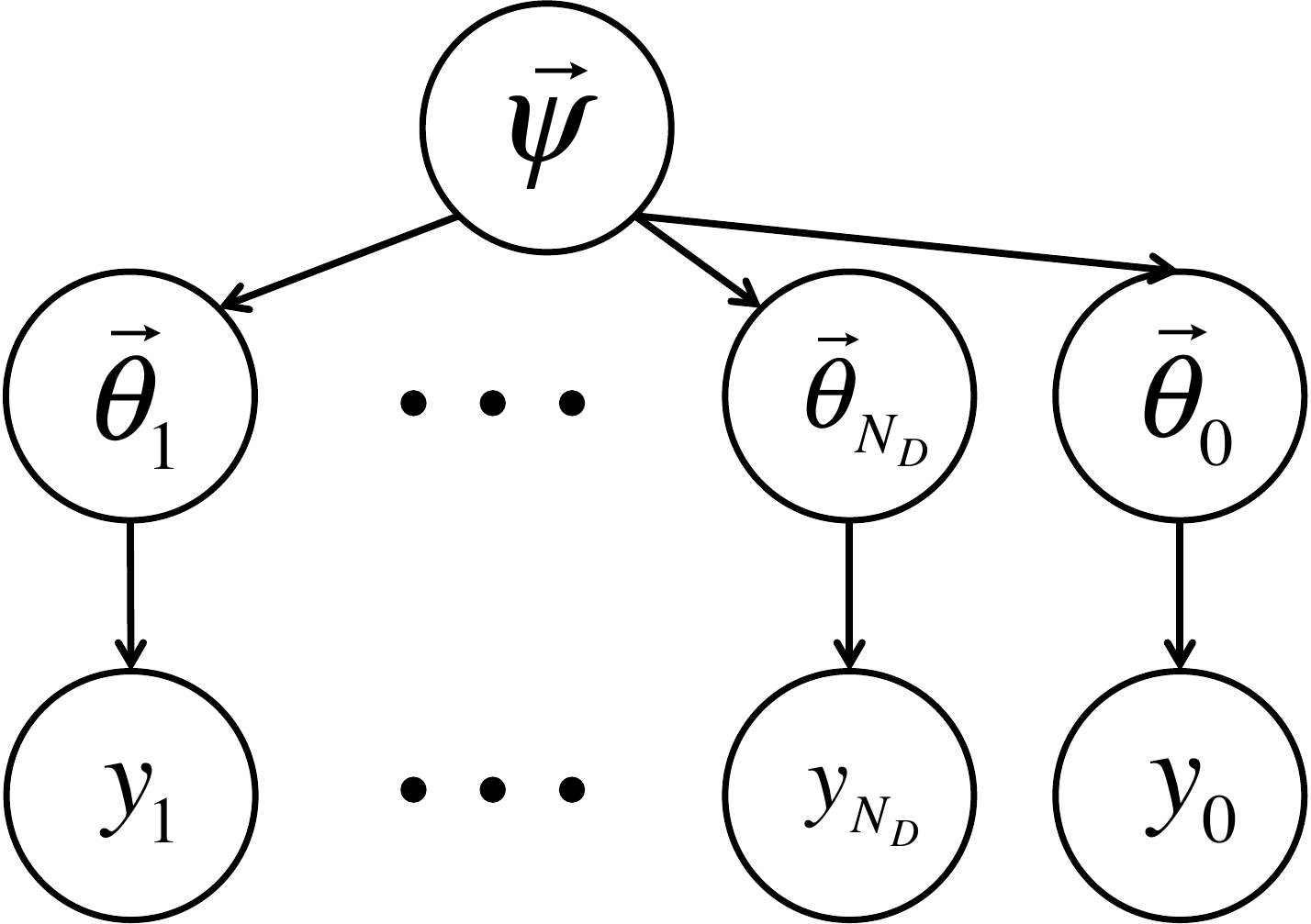}
		\caption{Hierarchical stochastic model}
		\label{fig:BN_HB}
	\end{subfigure}
	\caption{Graphical representations of the two HBM.}
	\label{fig:BN}
\end{figure}

HBM adds an extra level of hyperparameters $\vpsi$ to this classical Bayesian model. This new hierarchy can be added to either the prior (HPM) or the stochastic model/likelihood (HSM), which will result in completely different model classes suitable for different problems. The major distinction of the two models comes from the different structure of information dependencies between all the stochastic variables. The graph representations shown in Figure \ref{fig:BN} are very useful to display and study such relations \cite{Koller+Friedman:2009}. The arrows denote the directions of the forward models, which implies that Bayes' Theorem is needed for inference in the opposite direction. Any nodes without an incoming arrow requires a prior distribution. 

For HPM, an extra node is added to the starting nodes in order to parameterize the prior. This operation does not change the information dependency between the existing parameters. The posterior joint distribution of $\vtheta$ and $\vpsi$ is:
 \begin{equation}\label{eq:HPM_JP}
	p(\vtheta,\vpsi|\mD) = \frac{p(\mD|\vtheta)p(\vtheta|\vpsi)p(\vpsi)}{p(\mD)}.
\end{equation}
If we integrate out $\vpsi$ in the analysis, this is equivalent to recovering the classical Bayesian setting from HPM by choosing $p(\vtheta) = \int \! p(\vtheta|\vpsi)p(\vpsi)\,\mathrm{d}\vpsi$, because HPM does not affect the stochastic model part. Hence, the hyperparameters in HPM behave as latent variables. In most applications of HPM, we perform optimization of $\vpsi$ by maximizing $p(\mD|\vpsi)$, instead of the robust treatment of $\vpsi$. For example, in the context of SBL, zero-mean Gaussian priors with different variances are chosen for the coefficients in a linear regression problem. This prior is called the Automatic Relevance Determination (ARD) prior that is proven to induce sparsity \cite{Mackay:1994,Tipping:2004}. A commonly used algorithm in SBL, called the Relevance Vector Machine, is based on this optimization setup \cite{Tipping:2001,Tipping+Faul:2003}. One can interpret this procedure as a Bayesian model selection problem in the continuous space. 

For HSM, the hyperparameters affect the basic structure of the stochastic model. As shown in Figure \ref{fig:BN_HB}, each prediction is defined by a separate set of model parameters, and these parameters are correlated through the hyperparameters $\vpsi$. Here, the original model parameters behave as latent variables, and $\vpsi$ becomes the essential variables for future predictions $y_0$. The posterior distribution of $\vpsi$ is:
\begin{equation}\label{eq:HSM_PP}
	\begin{aligned}
		p(\vpsi|\mD) &= \frac{p(\mD|\vpsi)p(\vpsi)}{p(\mD)} \, \mathrm{d}\vpsi\\
		\text{where}~~ &p(\mD|\vpsi) = \prod_{i=1}^{N_D}p(D_i|\vpsi),\\
		\text{and}~~&p(D_i|\vpsi) = \int \! p(D_i|\vtheta_i)p(\vtheta_i|\vpsi) \, \mathrm{d}\vtheta_i.
	\end{aligned}
\end{equation}
Here, we denote the full data set as $\mD = \{D_i|i=1,\dots,N_D\}$, where $D_i$ represents the observed data values for $y_i$. The choice of $p(\vtheta_i|\vpsi)$ is flexible. For example, a statistical model is chosen in \cite{Gelman:2004bk} and \cite{Congdon:2010} and $\vpsi$ is called the hyperparameter vector. The statistical model can be seen as a common prior for all $\vtheta_i$. This explicit modeling for the prior separates different uncertainties by isolating out the uncertainty of $\vtheta_i$ across multiple groups of predictions. It can give better predictions and also be related to causal inference \cite{Gelman+Hill:2006}. Another example of such a hierarchical structure is found in a state-space model, where $\vtheta_i$ is the state variable and $p(\vtheta_i|\vpsi)$ represents the underlying stochastic process. Therefore, $p(\vtheta_i|\vpsi)$ can also be a theoretically informed model, instead of a purely empirical or statistical model.

In the current literatures, HPM is mainly used as a tool for optimal prior selection, while HSM is for explicit modeling of the uncertainty of model parameters across multiple groups of predictions for the data. However, HSM has the potential for more sophisticated modeling depending on the choice of $p(\vtheta_i|\vpsi)$, though it comes with a significantly larger computational cost due to the extra integral shown in Equation \ref{eq:HSM_PP}. We note that since HPM and HSM are two exclusive types of HBM, in theory, they can co-exist in a complex HBM and may have more than one level of hyperparameters. It is the limit of computational power that constraints the usage of HBM to be only one level of hypereparameters. In this paper, we study the important theoretical implications of HSM and present an efficient and flexible approximation method for performing Bayesian analysis with HSM. We note that a similar comparison can be found in \cite{Sargsyan:2015fr}, but they present it purely from the perspective of a likelihood function of $\vpsi$, instead of the HBM.

% Equations of posterior of psi
%\begin{equation}\label{eq:psi_HP}
%	\begin{aligned}
%		p(\vpsi|\mD,\mM_{HP}) &= \frac{p(\mD|\vpsi,\mM_{HP})p(\vpsi|\mM_{HP})}{p(\mD|\mM_{HP})}\\
%		&= \frac{\int \! \left[ \prod_{i=1}^{N_D} p(D_i|\vtheta,\mM_{HP})\right] p(\vtheta|\vpsi,\mM_{HP}) \, \mathrm{d} \vtheta ~ p(\vpsi|\mM_{HP})}{p(\mD|\mM_{HP})}
%	\end{aligned}
%\end{equation}
%\begin{equation}\label{eq:psi_HS}
%	\begin{aligned}
%	p(\vpsi|\mD,\mM_{HS}) &= \frac{\left[\prod_{i=1}^{N_D} p(D_i|\vpsi,\mM_{HS})\right]p(\vpsi|\mM_{HS})}{p(\mD|\mM_{HS})}\\
%&= \frac{\left[\prod_{i=1}^{N_D} \int \!  p(D_i|\vtheta_i,\mM_{HS}) p(\vtheta_i|\vpsi,\mM_{HS}) \, \mathrm{d} \vtheta_i \right] p(\vpsi|\mM_{HS})}{p(\mD|\mM_{HS})}
%\end{aligned}
%\end{equation}

\section{Theoretical Implications of HSM}\label{sec:theory}
Different assumptions on the uncertainties in a system can be represented using different hierarchical and non-hierarchical models. In this section, we verify that the effect of Occam's Razor in the Bayesian model selection framework is applicable to both types of models. Moreover, we study the theoretical implications of using a HSM in three important aspects:
\begin{enumerate}
	\item Separation of different types of uncertainties in a system (Section \ref{sec:2_setup})
	\item Identification of correlation between predictions or data (Section \ref{sec:3_R2})
	\item Uncertainty quantification of reduced order models (Section \ref{sec:3_R3})
\end{enumerate}
For computational accuracy, we demonstrate our results using polynomial regression because of the availability of many analytical solutions during the Bayesian analyses.

%\begin{equation}\label{eq:like1lv}
%\begin{aligned}
%p(\mD|\vec{\psi}) &= \int \! p(\mD | \vec{\theta}, \vec{\psi}) p(\vec{\theta}|\vec{\psi}) \, \mathrm{d}\vec{\theta}\\
%&= \int \! \left( \prod_{i=1}^{N_D} p(D_i | \vec{\theta},\vec{\psi}) \right) p(\vec{\theta}|\vec{\psi}) \, \mathrm{d}\vec{\theta}
%\end{aligned}
%\end{equation}
%\begin{equation}\label{eq:like2lv}
%\begin{aligned}
%p(\mD|\vec{\psi}) &= \prod_{i=1}^{N_D} p(D_i | \vec{\psi})\\
%&= \prod_{i=1}^{N_D} \int \!  p(D_i | \vec{\theta}_i,\vec{\psi}) p(\vec{\theta}_i|\vec{\psi}) \, \mathrm{d} \vec{\theta}_i
%\end{aligned}
%\end{equation}

\subsection{Separating different types of uncertainties using HSM}\label{sec:2_setup}
\begin{figure}[h]
	\centering
	\includegraphics[width=\textwidth]{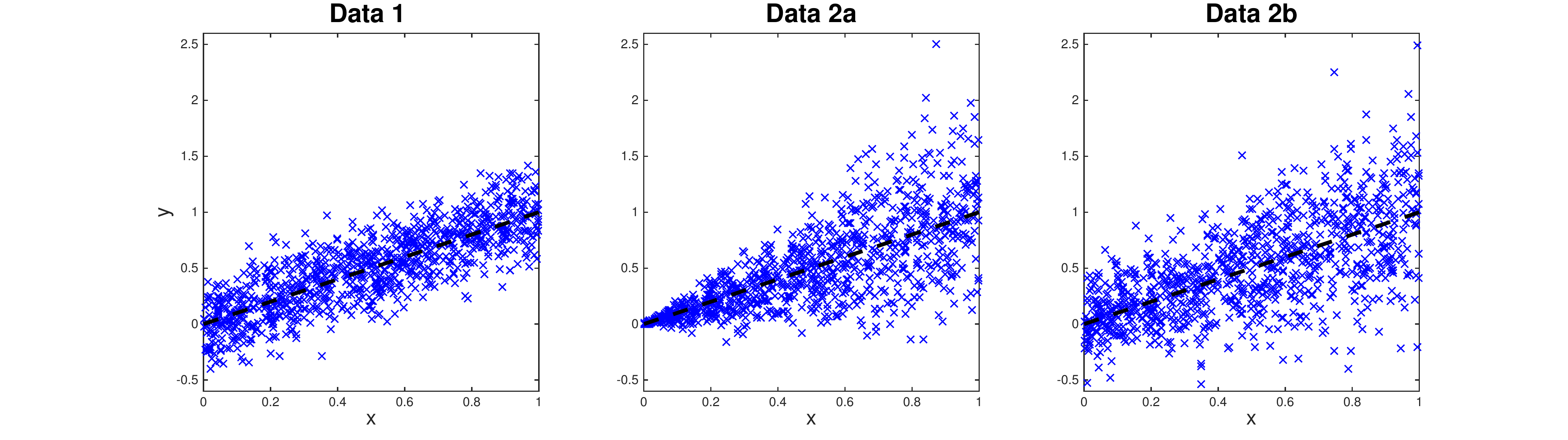}
	\caption{Three different types of contaminated data sets with $x$ between 0 and 1. The black dash lines denote the actual function without any error.}
	\label{fig:C1_d1}
\end{figure}

\begin{figure}[h]
	\centering
	\includegraphics[width=\textwidth]{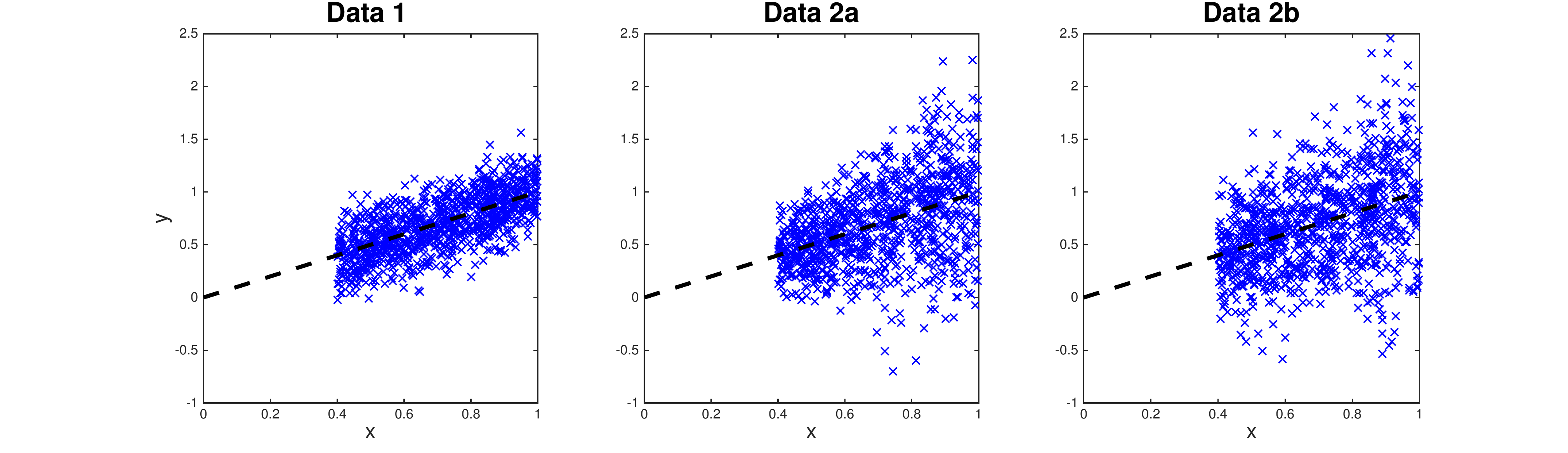}
	\caption{Three different types of contaminated data sets with $x$ between 0.4 and 1. The black dash lines denote the actual function without any error.}
	\label{fig:C1_d2}
\end{figure}

In this study, we demonstrate how HSM can capture different types of uncertainties using models that are embedded with different sources of stochasticity. We generate synthetic data based on two uncertainty models: (1) zero-mean Gaussian error $\epsilon_y$ added to the function $f(x,\vec{\theta})$, which can represent measurement noise in practice; (2) zero-mean Gaussian error $\epsilon_\theta$ added to the model parameters $\vec{\theta}$, which can represent inherent model uncertainty due to, for example, insufficient knowledge of theoretically informed models or environmental variations across multiple experiments. We use a linear function $f(x,\theta) = \theta x$ ($\theta$ and $x$ are scalars) and all data points are generated with independent $\epsilon_y$ and $\epsilon_\theta$. Three types of synthetic data are considered based on $\hat{\theta}$, a fixed value of the model parameter $\theta$:
\begin{enumerate}
	\item Additive error data, $\mD_1$: $y = f(x,\hat{\theta}) + \epsilon_y, \,\,\, \epsilon_y \sim N(\epsilon_y|0,\hat{\sigma}_y^2)$
	\item Embedded error data, $\mD_{2a}$: $y = f(x,\hat{\theta} + \epsilon_\theta), \,\,\, \epsilon_\theta \sim N(\epsilon_\theta|0,\hat{\sigma}_\theta^2)$
	\item Mixed error data, $\mD_{2b}$: $y = f(x,\hat{\theta} + \epsilon_\theta) + \epsilon_y, \,\,\, \epsilon_y \sim N(\epsilon_y|0,\hat{\sigma}_y^2)$ and $\epsilon_\theta \sim N(\epsilon_\theta|0,\hat{\sigma}_\theta^2)$
\end{enumerate}
where $N(z|\mu,\sigma^2)$ denotes a Gaussian distribution for $z$ with mean $\mu$ and variance $\sigma^2$. We set $\hat{\theta} = 1$, $\hat{\sigma}_\theta = 0.5$ and $\hat{\sigma}_y = 0.2$. Each data set contains 1000 data points independently generated from a uniformly distributed $x$ value between preset bounds and random errors $\epsilon_y$ and $\epsilon_\theta$ from the corresponding Gaussian distributions.  We generate all three types of data twice, once with $x$ between 0 and 1 and once with $x$ between 0.4 and 1 (see Figures \ref{fig:C1_d1} and \ref{fig:C1_d2}, respectively).

Given the three types of data, we perform Bayesian model class selection on four different stochastic model classes:
\begin{enumerate}
	\item Non-hierarchical model, $\mM_{1a}$: This model class includes a uniform prior for $\theta$ between -1 and 3, and a Gaussian likelihood for $\theta$, i.e., $p(\{x_i,y_i\}|\theta,\sigma_y) = N(y_i|f(x_i,\theta),\sigma_y^2)$ for any $i$. Here, $\sigma_y$ is also treated as a parameter to be inferred. We use a uniform prior for $\sigma_y$ between 0.001 and 1.
	\item HPM, $\mM_{1b}$: This model class has the same likelihood as $\mM_{1a}$ and a hierarchical prior for $\theta$. The prior is distributed as $N(\theta|\mu_\theta,\sigma_\theta^2)$, i.e., the hyperparameters $\vpsi = \{\mu_\theta,\sigma_\theta\}$. We use uniform priors for $\mu_\theta$ between -1 and 3 and $\sigma_\theta$ between 0.001 and 1. The prior for $\sigma_y$ follows the one in $\mM_{1a}$.
	\item Zero noise HSM, $\mM_{2a}$: This is an HSM with $p(\theta_i|\vpsi)$ modeled as the same Gaussian prior used for $\theta$ in $\mM_{1b}$ for all $i$. No additive error is assumed in this model. This implies that the likelihood function of $\theta_i$ is a delta function centered at the given data, i.e., $p(\{x_i,y_i\}|\theta_i) = \delta(y_i-f(x_i,\theta_i))$. The priors of the hyperparameters are the same as in $\mM_{1b}$.
	\item Full HSM, $\mM_{2b}$: This is the same HSM as $\mM_{2a}$ except that a Gaussian additive error is assumed in this model. Hence, the likelihood function $p(\{x_i,y_i\}|\theta_i,\sigma_y) = N(y_i|f(x_i,\theta_i),\sigma_y^2)$. We use the same uniform priors as in $\mM_{1b}$ for $\sigma_y$, $\mu_\theta$ and $\sigma_\theta$.
\end{enumerate}

Appendix \ref{app:A} includes all derivations of the analytical expressions used for estimating the evidence $p(\mD|\mM_k)$, the marginal posterior probability density function (PDF) for the model parameters $p(\theta|\mD,\mM_k)$, and the posterior PDF for the hyperparameters $p(\psi|\mD,\mM_k)$, for a given model class $\mM_k$. We use Monte Carlo Simulation to draw posterior samples of the hyperparameters for the ``naive" estimation of the evidence. 10K samples are used to obtain accurate estimates with small variances. 

\subsubsection{Results and discussion}\label{sec:3_R1}

Table \ref{tab:C1a} and \ref{tab:C1b} summarize the results of the Bayesian inference and model selection for all six data sets (three with $x$ between 0 and 1, and three with $x$ between 0.4 and 1). The Bayesian model selection framework is capable of selecting the corresponding model used to generate the data. All models have a relatively good estimate for $\hat{\theta}$ given by $E[\theta|\mD]$. The major difference appears in the accuracy of estimating $\hat{\sigma}_\theta$ and $\hat{\sigma}_y$ given by $Std[\theta|\mD]$ and $E[\sigma_y|\mD]$, respectively. In general, in terms of accurately estimating $\hat{\sigma}_\theta$ and $\hat{\sigma}_y$, $\mM_{1a}$ and $\mM_{1b}$ always prefer putting uncertainty into $\sigma_y$ and thus perform well for $\mD_1$ only. $\mM_{2a}$ cannot handle additive error and thus performs well for $\mD_{2a}$ only. $\mM_{2b}$ is the most flexible model and it performs relatively well for all data sets $\mD_1$, $\mD_{2a}$ and $\mD_{2b}$.

We note that the results of $\mM_{1a}$ and $\mM_{1b}$ are extremely similar. This is because their only difference is the prior of $\theta$, as explained in Section \ref{sec:HBM}. In this case, the marginalized prior $p(\theta)$ in $\mM_{1b}$ is slightly larger than the prior in $\mM_{1a}$. Hence, the evidence of both models are almost the same in all cases.

Furthermore, we observe that in the case of $\mD_{2b}$ with $x$ between 0.4 and 1, $\mM_{2a}$ and $\mM_{2b}$ have a similar posterior model probability. Intuitively, the data with small $x$ values are more sensitive to additive noise because the noise-to-signal ratio is much higher than in the case of larger $x$ values. When such data is not available, it is challenging to distinguish between $\mD_{2a}$ and $\mD_{2b}$. This can be proved visually by observing the similarity between $\mD_{2a}$ and $\mD_{2b}$ in Figure \ref{fig:C1_d2} as compared to the one in Figure \ref{fig:C1_d1}.

%We did not experience the degenerate problem observed in \cite{Sargsyan:2015kq} because the analytical expressions resolve the issue of sampling with a delta likelihood function. Hence, this is only a computational issue when the analytical expressions are not used. 
Our results suggest using $\mM_{2b}$ for model calibration, when no prior knowledge indicates that additive error is irrelevant. This HSM can appropriately separates the two types of uncertainties in our case study into the additive and non-additive parts based on the observed data. If computational power allows, Bayesian model class selection should always be performed among a set of different candidate models, as illustrated in this study.

\begin{table}
	\footnotesize
	\caption{Results for testing different stochastic model classes on the linear function (input $x$ between 0 and 1). The row labeled ``Ref." shows the actual values for the corresponding estimates.}
	\label{tab:C1a}
	\centering
	\begin{tabular}{|c|c|c|c|c|c|c|c|}
		\hline
		Data & Model & $E[\theta|\mD]$ & $Std[\theta|\mD]$ & $E[\sigma_y|\mD]$ & $Std[\sigma_y|\mD]$ & ln(Evid.) & $P(\mM'_k|\mD)$ \\ \hline
		\hline
		\multirow{5}{*}{$\mD_1$} & $\mM_{1a}$ & 1.012 & 0.011 & 0.2030 & 0.0046 & 167.0 & 0.570\\ \cline{2-8}
		& $\mM_{1b}$ & 1.012 & 0.011 & 0.2033 & 0.0049 & 166.7 & 0.424\\ \cline{2-8}
		& $\mM_{2a}$ & 0.996 & 2.000 & 0.0000 & 0.0000 & -2536.7 & 0.000\\ \cline{2-8}
		& $\mM_{2b}$ & 1.017 & 0.133 & 0.1796 & 0.0008 & 162.4 & 0.006\\ \cline{2-8}
		& Ref. & 1.000 & 0.000 & 0.200 & - - - & - - - & - - - \\ \hline
		\hline
		\multirow{5}{*}{$\mD_{2a}$} & $\mM_{1a}$ & 1.002 & 0.016 & 0.2808 & 0.0065 & -158.2 & 0.000\\ \cline{2-8}
		& $\mM_{1b}$ & 1.002 & 0.016 & 0.2814 & 0.0067 & -158.4 & 0.000\\ \cline{2-8}
		& $\mM_{2a}$ & 0.978 & 0.525 & 0.0000 & 0.0000 & 285.3 & 1.000\\ \cline{2-8}
		& $\mM_{2b}$ & 1.042 & 0.572 & 0.0021 & 0.0000 & 264.6 & 0.000\\ \cline{2-8}
		& Ref. & 1.000 & 0.500 & 0.000 & - - - & - - - & - - - \\ \hline
		\hline
		\multirow{5}{*}{$\mD_{2b}$} & $\mM_{1a}$ & 0.998 & 0.019 & 0.3334 & 0.0073 & -328.6 & 0.000\\ \cline{2-8}
		& $\mM_{1b}$ & 0.998 & 0.019 & 0.3330 & 0.0071 & -328.7 & 0.000\\ \cline{2-8}
		& $\mM_{2a}$ & 0.967 & 2.001 & 0.0000 & 0.0000 & -16824.3 & 0.000\\ \cline{2-8}
		& $\mM_{2b}$ & 1.027 & 0.453 & 0.1956 & 0.0041 & -260.6 & 1.000\\ \cline{2-8}
		& Ref. & 1.000 & 0.500 & 0.200 & - - - & - - - & - - - \\ \hline
	\end{tabular}
\end{table}

\begin{table}
	\footnotesize
	\caption{Results for testing different stochastic model classes on the linear function (input $x$ between 0.4 and 1). The row labeled ``Ref." shows the actual values for the corresponding estimates.}
	\label{tab:C1b}
	\centering
	\begin{tabular}{|c|c|c|c|c|c|c|c|}
		\hline
		Data & Model & $E[\theta|\mD]$ & $Std[\theta|\mD]$ & $E[\sigma_y|\mD]$ & $Std[\sigma_y|\mD]$ & ln(Evid.) & $P(\mM'_k|\mD)$ \\ \hline
		\hline
		\multirow{5}{*}{$\mD_1$} & $\mM_{1a}$ & 0.993 & 0.009 & 0.1985 & 0.0044 & 190.5 & 0.498\\ \cline{2-8}
		& $\mM_{1b}$ & 0.993 & 0.009 & 0.1986 & 0.0050 & 190.5 & 0.486\\ \cline{2-8}
		& $\mM_{2a}$ & 0.994 & 0.313 & 0.0000 & 0.0000 & 121.8 & 0.000\\ \cline{2-8}
		& $\mM_{2b}$ & 0.984 & 0.035 & 0.2086 & 0.0017 & 187.0 & 0.015\\ \cline{2-8}
		& Ref. & 1.000 & 0.000 & 0.200 & - - - & - - - & - - - \\ \hline
		\hline
		\multirow{5}{*}{$\mD_{2a}$} & $\mM_{1a}$ & 0.998 & 0.016 & 0.3659 & 0.0082 & -421.1 & 0.000\\ \cline{2-8}
		& $\mM_{1b}$ & 0.997 & 0.016 & 0.3666 & 0.0072 & -421.0 & 0.000\\ \cline{2-8}
		& $\mM_{2a}$ & 1.000 & 0.500 & 0.0000 & 0.0000 & -342.0 & 0.986\\ \cline{2-8}
		& $\mM_{2b}$ & 1.006 & 0.469 & 0.0141 & 0.0001 & -346.2 & 0.014\\ \cline{2-8}
		& Ref. & 1.000 & 0.500 & 0.000 & - - - & - - - & - - - \\ \hline
		\hline
		\multirow{5}{*}{$\mD_{2b}$} & $\mM_{1a}$ & 0.980 & 0.018 & 0.4136 & 0.0095 & -542.4 & 0.000\\ \cline{2-8}
		& $\mM_{1b}$ & 0.980 & 0.018 & 0.4136 & 0.0096 & -542.7 & 0.000\\ \cline{2-8}
		& $\mM_{2a}$ & 0.984 & 0.596 & 0.0000 & 0.0000 & -508.9 & 0.558\\ \cline{2-8}
		& $\mM_{2b}$ & 0.946 & 0.500 & 0.2101 & 0.0182 & -509.1 & 0.442\\ \cline{2-8}
		& Ref. & 1.000 & 0.500 & 0.200 & - - - & - - - & - - - \\ \hline
	\end{tabular}
\end{table}

\subsubsection{Effect of grouping}\label{sec:3_R2}
In the previous section, all synthetic data is generated independently based on different stochastic models. In practice, we may group some of the predictions to be modeled under one set of parameter values $\vtheta_i$, where $i$ is the index for each group of predictions. For example, if an experiment is repeated in different laboratories, we may model the data from the same laboratory using the same set of model parameters, assuming the data shares the same environmental factors during the experiment. When the prediction grouping is not known, we may want to know the most plausible grouping, which defines a unique HSM. 

\begin{figure}
\centering
\begin{subfigure}[t]{0.4\textwidth}
	\centering
	\includegraphics[width=\textwidth]{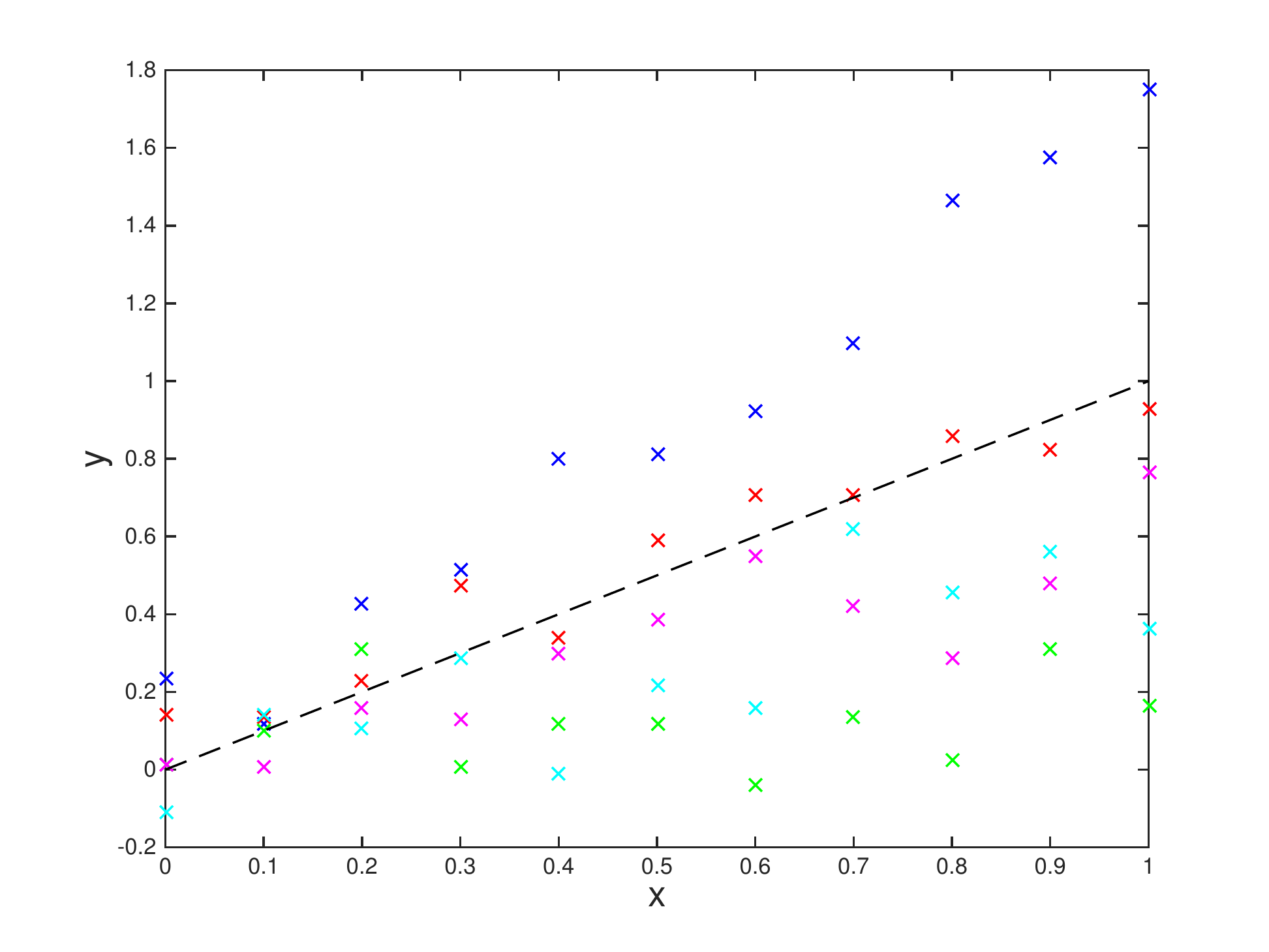}
	\caption{Actual grouping}
	\label{fig:C2_d1}
\end{subfigure}
\qquad
\begin{subfigure}[t]{0.4\textwidth}
	\centering
	\includegraphics[width=\textwidth]{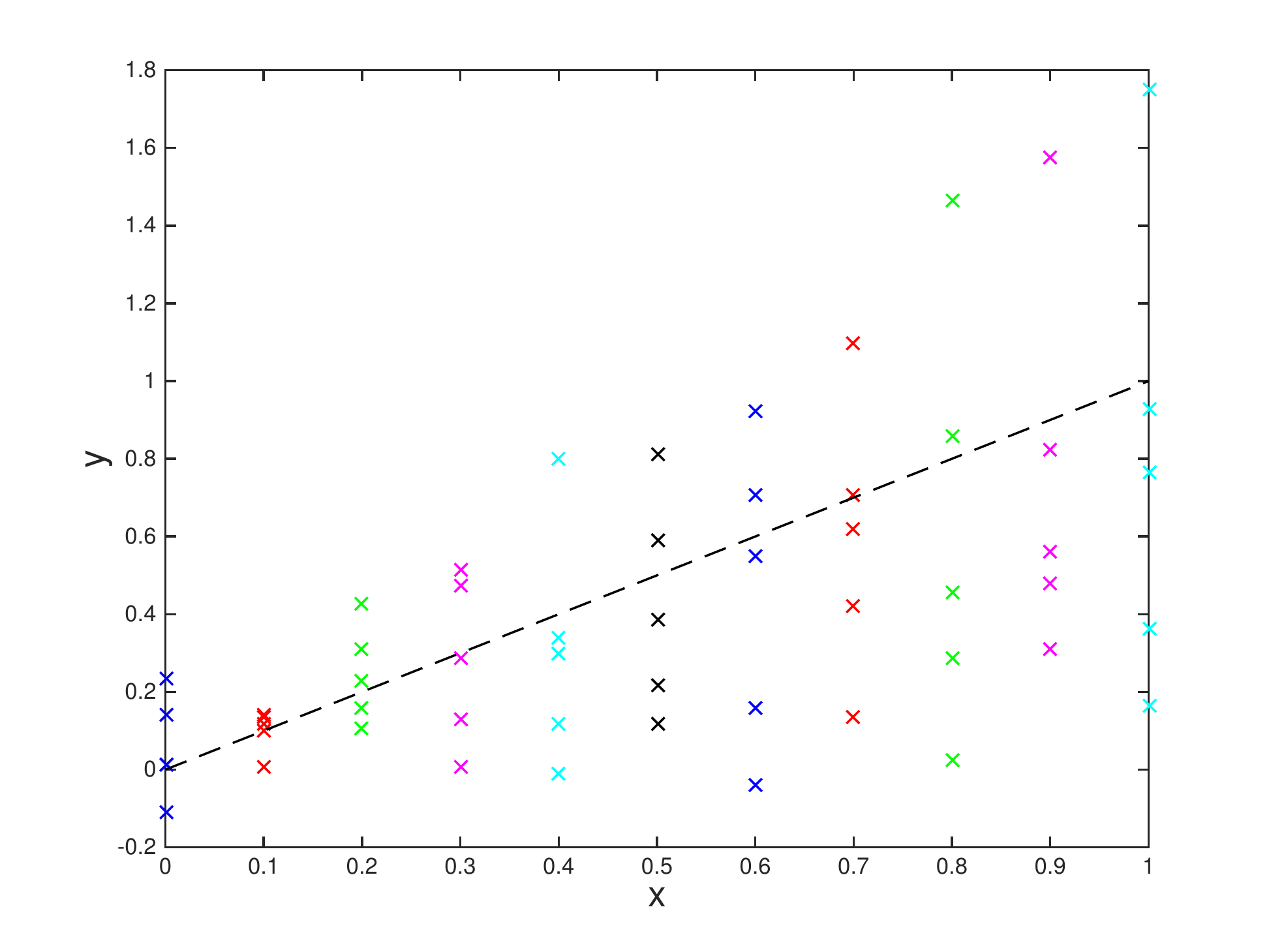}
	\caption{Grouping across $x$}
	\label{fig:C2_d2}
\end{subfigure}
\qquad
\begin{subfigure}[t]{0.4\textwidth}
	\centering
	\includegraphics[width=\textwidth]{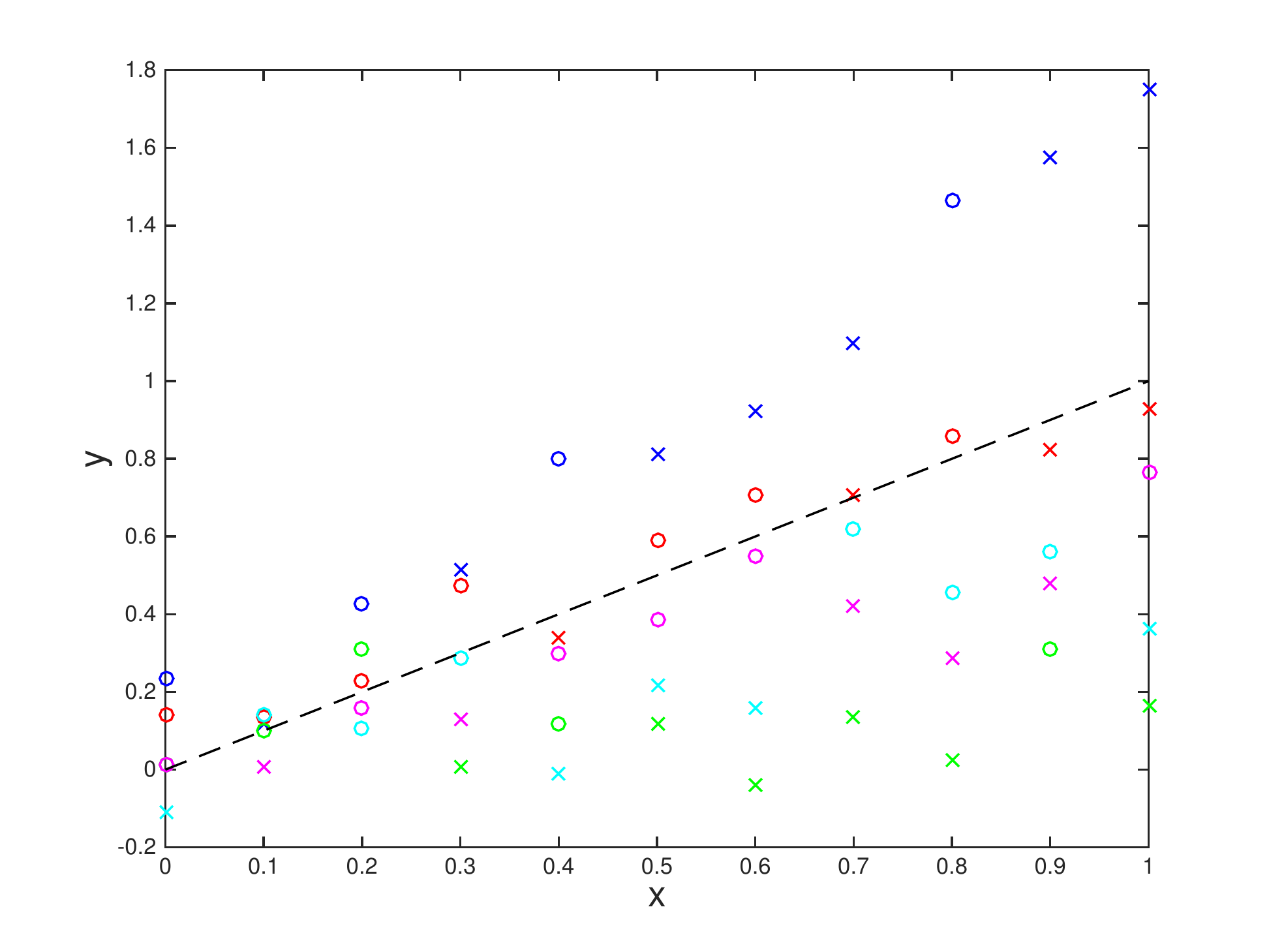}
	\caption{Over grouping}
	\label{fig:C2_d3}
\end{subfigure}
\qquad
\begin{subfigure}[t]{0.4\textwidth}
	\centering
	\includegraphics[width=\textwidth]{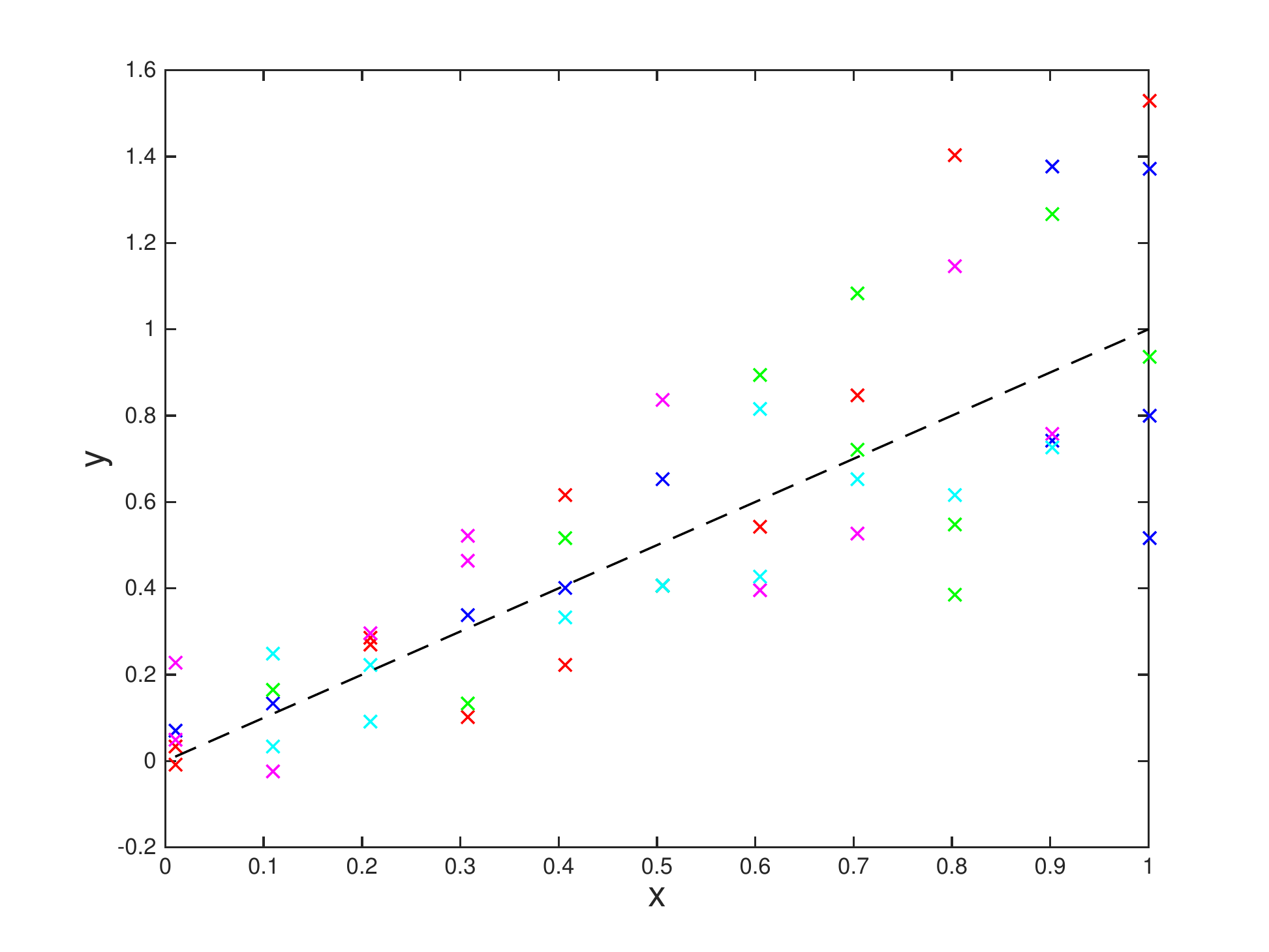}
	\caption{Random grouping}
	\label{fig:C2_d6}
\end{subfigure}
\caption{Different groupings of data sets. Different colors and marker combinations represent different data sets. The dashed lines denote the actual function without any error.}
\label{fig:Group}
\end{figure}

We use the same linear function $f(x,\theta) = \theta x$ as before, but generate a new set of data $\mD$. First of all, five random samples $\theta^{(i)}$ for $i = 1,\dots,5$ are drawn from $N(\theta|\hat{\mu}_\theta,\hat{\sigma}_\theta^2)$. Then, 11 data points are generated for each $\theta^{(i)}$ to form a data set $D_i$ by using the stochastic forward model $y = f(x,\theta^{(i)}) + \epsilon_y$, where $\epsilon_y$ is random error drawn from $N(\epsilon_y|0,\hat{\sigma}_y^2)$ independently for each data point. In this study, $\hat{\mu}_\theta = 1$, $\hat{\sigma}_\theta = 0.5$ and $\hat{\sigma}_y = 0.1$. Figure \ref{fig:C2_d1} shows the five data sets used in the section. The sample mean and standard deviation of $\theta^{(i)}$ are 1.0191 and 0.4079, respectively. 

To study the effect of grouping predictions in the HSM, we set up six different HSMs based on different information dependency between the predictions. We perform Bayesian model class selection to find the most plausible HSM to describe the data. The evidence is calculated similarly to $\mM_{2b}$ in Section \ref{sec:2_setup}. Details of the calculation can be found in Appendix \ref{app:A}. The six data groupings are:
\begin{enumerate}
	\item Actual grouping, $\mM'_1$: group predictions according to Figure \ref{fig:C2_d1}.
	\item Constant $x$ grouping, $\mM'_2$: predictions with the same $x$ value are grouped together. This case represents grouping data according to the wrong variable. Figure \ref{fig:C2_d2} shows the resulting grouping.
	\item 1/2 error grouping, $\mM'_3$: starting from $\mM'_1$, the predictions for each data set are further divided into 2 sets by the center line of the data points in the set. This center line is defined by the mean of $\theta$ inferred from each data point in the set. Figure \ref{fig:C2_d3} shows the resulting data grouping.
	\item 1/4 error grouping, $\mM'_4$: starting from $\mM'_3$, the predictions for each data set are again further divided into 2 sets by the center line of the data points in the set, as done in $\mM'_3$.
	\item Single prediction, $\mM'_5$: each prediction is treated as an independent data set (same as Section \ref{sec:3_R1}).
	\item Random grouping, $\mM'_6$: predictions for the data points are randomly collected into five groups. Figure \ref{fig:C2_d6} shows the resulting grouping.
\end{enumerate}

%\begin{figure}
%	\centering
%	%\input{figures/fig_Data}
%	\includegraphics[width=0.5 \textwidth]{figures/C2_d1.eps}
%	\caption{Five sets of data with different $\theta$ values. Each color represents one data set. The dash line denotes the actual function without any error.}
%	\label{fig:C2_d0}
%\end{figure}
%
%\begin{figure}
%	\centering
%	%\input{figures/fig_Data}
%	\includegraphics[width=0.5 \textwidth]{figures/C2_d2.eps}
%	\caption{Eleven sets of data, each with the same $x$ values. The color is for visualization. Same color does not necessarily represent the same data set.}
%	\label{fig:C2_d2}
%\end{figure}
%
%\begin{figure}
%	\centering
%	%\input{figures/fig_Data}
%	\includegraphics[width=0.5 \textwidth]{figures/C2_d3.eps}
%	\caption{Ten sets of data divided based on $\mM'_1$. Different color and marker combination represents different data set.}
%	\label{fig:C2_d3}
%\end{figure}
%
%\begin{figure}
%	\centering
%	%\input{figures/fig_Data}
%	\includegraphics[width=0.5 \textwidth]{figures/C2_d6.eps}
%	\caption{Five sets of randomly grouped data. Each color represents one data set.}
%	\label{fig:C2_d6}
%\end{figure}

Table \ref{tab:C2} summarizes the results of the Bayesian inference and model class selection for all six HSM. All models have a good estimate for $\hat{\mu}_\theta$ (corresponds to $E[\theta|\mD]$). However, $\mM'_2$ and $\mM'_6$, representing two completely wrong groupings, have significantly lower estimates for $\hat{\sigma}_\theta$ (given by $Std[\theta|\mD]$), but larger estimates for the mean of $\hat{\sigma}_y$ (given by $E[\sigma_y|\mD]$). Hence, they have a very low log-evidence value. $\mM'_1$ has the closest estimates of $\hat{\mu}_\theta$, $\hat{\sigma}_\theta$ and $\hat{\sigma}_y$, but it is not the most probable model class among the six. $\mM'_3$ is the most probable model class even though it has a low value of the estimates for both $\hat{\sigma}_\theta$ and $\hat{\sigma}_y$. We note that when the total number of data points is fixed, the more groups there are, the smaller the average number of predictions in a group is. This affects the likelihood of $\vpsi$, which equals the product of $p(D_i|\vpsi)$ for each data set $D_i$ (see Equation \ref{eq:HSM_PP}). When the number of $p(D_i|\vpsi)$ increases, the value of each evidence term may decrease. This decrease is due to a less peaked $p(D_i|\vtheta_i)$ as the number of data points in the set decreases. The final evidence $p(\mD|\mM'_k)$ for a given data grouping $\mM'_k$ is a tradeoff between these two factors. Starting from $\mM'_1$ being modified to $\mM'_3$, if we repeat the process many times, eventually we will reach $\mM'_5$. Therefore, $\mM'_1$, $\mM'_3$, $\mM'_4$ and $\mM'_5$ can be considered as a sequence of similar data grouping methods. In the end, the tradeoff suggests that $\mM'_3$ is the most probable  grouping. This implies that model class selection based on the evidence may not necessarily lead to the actual grouping, if it exists. Overall, it tries to minimize uncertainties in all parameters. On the other hand, if we choose the actual grouping, the uncertainty quantification will be accurate.

\begin{table}
	\footnotesize
	\caption{Results for testing different HSM.}
	\label{tab:C2}
	\centering
	\begin{tabular}{|c|c|c|c|c|c|c|}
		\hline
		& $E[\theta|\mD]$ & $Std[\theta|\mD]$ & $E[\sigma_y|\mD]$ & $Std[\sigma_y|\mD]$ & ln(Evid.) & $P(\mM'_k|D)$ \\ \hline
		$\mM'_1$ & 1.050 & 0.528 & 0.091 & 0.009 & 36.18 & 0.0012\\ \hline
		$\mM'_2$ & 1.051 & 0.061 & 0.230 & 0.023 & -3.42 & 0.0000\\ \hline
		$\mM'_3$ & 1.036 & 0.473 & 0.066 & 0.007 & 42.88 & 0.9986\\ \hline
		$\mM'_4$ & 1.024 & 0.432 & 0.061 & 0.008 & 33.91 & 0.0001\\ \hline
		$\mM'_5$ & 1.052 & 0.348 & 0.097 & 0.020 & 5.49 & 0.0000\\ \hline
		$\mM'_6$ & 1.053 & 0.112 & 0.228 & 0.023 & -3.14 & 0.0000\\ \hline
		\hline
		& ($\hat{\mu}_\theta$) & ($\hat{\sigma}_\theta$) & ($\hat{\sigma}_y$) & & & \\ \hline
		$Ref.$ & 1.000 & 0.500 & 0.100 & - - - & - - - & - - -\\ \hline
	\end{tabular}
\end{table}

\subsection{Uncertainty quantification for reduced order models using HSM}\label{sec:3_R3}
A reduced order model simplifies the information obtained in the original model. This loss of information can be treated as a source of uncertainty in the Bayesian framework. The structure of this type of uncertainty can be defined if the mapping between the original model and the reduced order model is completely known. In most cases, however, the mapping is not well-defined or too complicated to be studied directly. Therefore, there is nono clear way to model such kind of loss of information. In this section, we compare the performance of using different model classes in Section \ref{sec:2_setup} to represent the reduced order model uncertainty. This study is based on fitting second and third degree polynomial data with linear functions.

\subsubsection{Problem setup for polynomials}
\begin{figure}
	\centering
	\includegraphics[width=\textwidth]{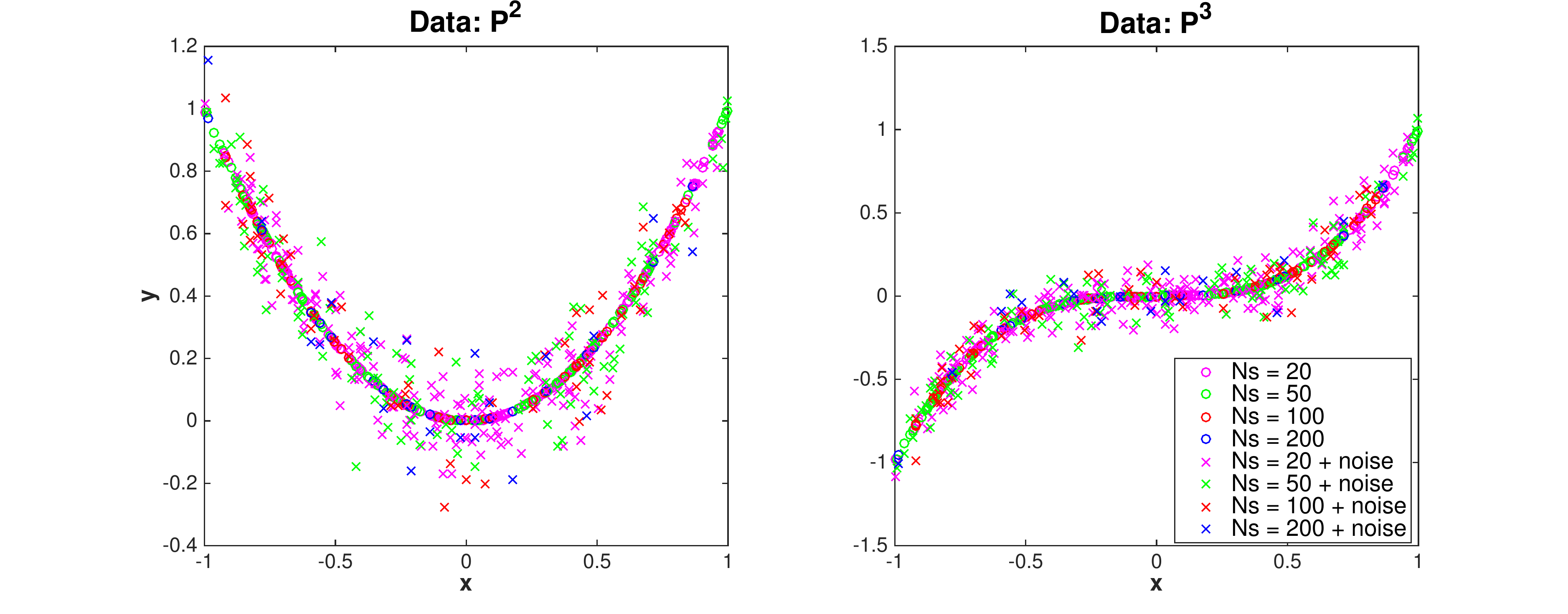}
	\caption{8 data sets from $y = x^2$ ($P^2$) and 8 data sets from $y = x^3$ ($P^3$). The noises are additive Gaussian error with standard deviation $\hat{\sigma}_y = 0.1$.}
	\label{fig:C3_raw}
\end{figure}

We consider a total of 16 sets of data: 8 sets from a quadratic function $f(x) = x^2$ and 8 sets from a cubic function $f(x) = x^3$. For each type of function, we generate 20, 50, 100 and 200 data points twice (with and without noise), i.e., a total of 8 data sets. The $x$ values of the data points are generated randomly in the interval $[-1,1]$. Additive Gaussian error with standard deviation $\hat{\sigma}_y = 0.1$ is chosen for the noise. Figure \ref{fig:C3_raw} shows all 16 data sets used in this study. 

We perform Bayesian model class selection and posterior robust prediction using three models described in Section \ref{sec:2_setup}: $\mM_{1a}$ (denoted as $\mM_1$ in this section), $\mM_{2a}$ and $\mM_{2b}$. Appendix \ref{app:A} includes all derivations for the analytical expressions used to estimate the log-evidence values and posterior robust predictions $p(y|\mD,\mM_k)$ for a given model $\mM_k$. We evaluate the posterior probability of observing each grid point $(\hat{x},\hat{y})$ on a 2D fine grid of $(x,y)$ using the analytical expressions in order to construct the distribution of the posterior robust prediction.

\subsubsection{Results and discussion}
Table \ref{tab:C3} summarizes the results of model class selection for this study. $\mM_{2b}$ is the most probable model class for data with additive noise and $\mM_{2a}$ is the most probable model class for data without any additive noise. $\mM_1$ is the least significant model in most cases because the error caused by the reduced order model is very non-linear. We do not expect that the additive noise alone is sufficient to explain the error. In this study, the sample size (number of data points) does not play a significant role in the results. 

\begin{table}
	\footnotesize
	\caption{Model class selection results for studying uncertainty quantification of reduced order models.}
	\label{tab:C3}
	\centering
	\begin{tabular}{|c|c|c|c|c|c|c|c|c|c|}
		\hline
		Sample & & \multicolumn{2}{|c|}{Quadratic (no noise)} & \multicolumn{2}{c|}{Quadratic (with noise)} & \multicolumn{2}{c|}{Cubic (no noise)} & \multicolumn{2}{c|}{Cubic (with noise)}\\ \cline{3-10}
		Size & & ln(Evid.) & $P(\mM_k|\mD)$ & ln(Evid.) & $P(\mM_k|\mD)$ & ln(Evid.) & $P(\mM_k|\mD)$ & ln(Evid.) & $P(\mM_k|\mD)$\\ \hline \hline
		\multirow{3}{*}{20} & $\mM_1$ & -10.1 & 0.0000 & -11.4 & 0.0095 & 5.9 & 0.0000 & 0.4 & 0.0908\\ \cline{2-10}
		& $\mM_{2a}$ & 8.6 & 0.9873 & -15.4 & 0.0002 & 18.4 & 0.9773 & -8.9 & 0.0000\\ \cline{2-10}
		& $\mM_{2b}$ & 4.3 & 0.0127 & -6.8 & 0.9903 & 14.7 & 0.0227 & 2.7 & 0.9092\\ \hline \hline
		\multirow{3}{*}{50} & $\mM_1$ & -31.2 & 0.0000 & -33.4 & 0.0018 & 28.5 & 0.0035 & 12.1 & 0.9625\\ \cline{2-10}
		& $\mM_{2a}$ & -6.2 & 0.9920 & -1014.3 & 0.0000 & 34.2 & 0.9927 & -237.2 & 0.0000\\ \cline{2-10}
		& $\mM_{2b}$ & -11.0 & 0.0080 & -27.1 & 0.9982 & 28.6 & 0.0037 & 8.9 & 0.0375\\ \hline \hline 
		\multirow{3}{*}{100} & $\mM_1$ & -67.1 & 0.0000 & -66.2 & 0.0000 & 35.9 & 0.0000 & 21.3 & 0.0020\\ \cline{2-10}
		& $\mM_{2a}$ & -8.6 & 0.9685 & -61.4 & 0.0000 & 60.8 & 0.9989 & -31.2 & 0.0000\\ \cline{2-10}
		& $\mM_{2b}$ & -12.0 & 0.0315 & -34.1 & 1.0000 & 53.9 & 0.0011 & 27.5 & 0.9980\\ \hline \hline
		\multirow{3}{*}{200} & $\mM_1$ & -101.9 & 0.0000 & -102.3 & 0.0000 & 117.9 & 0.0000 & 80.6 & 0.0000\\ \cline{2-10}
		& $\mM_{2a}$ & 57.7 & 0.9864 & -272.5 & 0.0000 & 194.9 & 1.0000 & -559.6 & 0.0000\\ \cline{2-10}
		& $\mM_{2b}$ & 53.4 & 0.0136 & -10.3 & 1.0000 & 170.0 & 0.0000 & 92.7 & 1.0000\\ \hline
	\end{tabular}
\end{table}

\begin{figure}
	\centering
	\includegraphics[width=\textwidth]{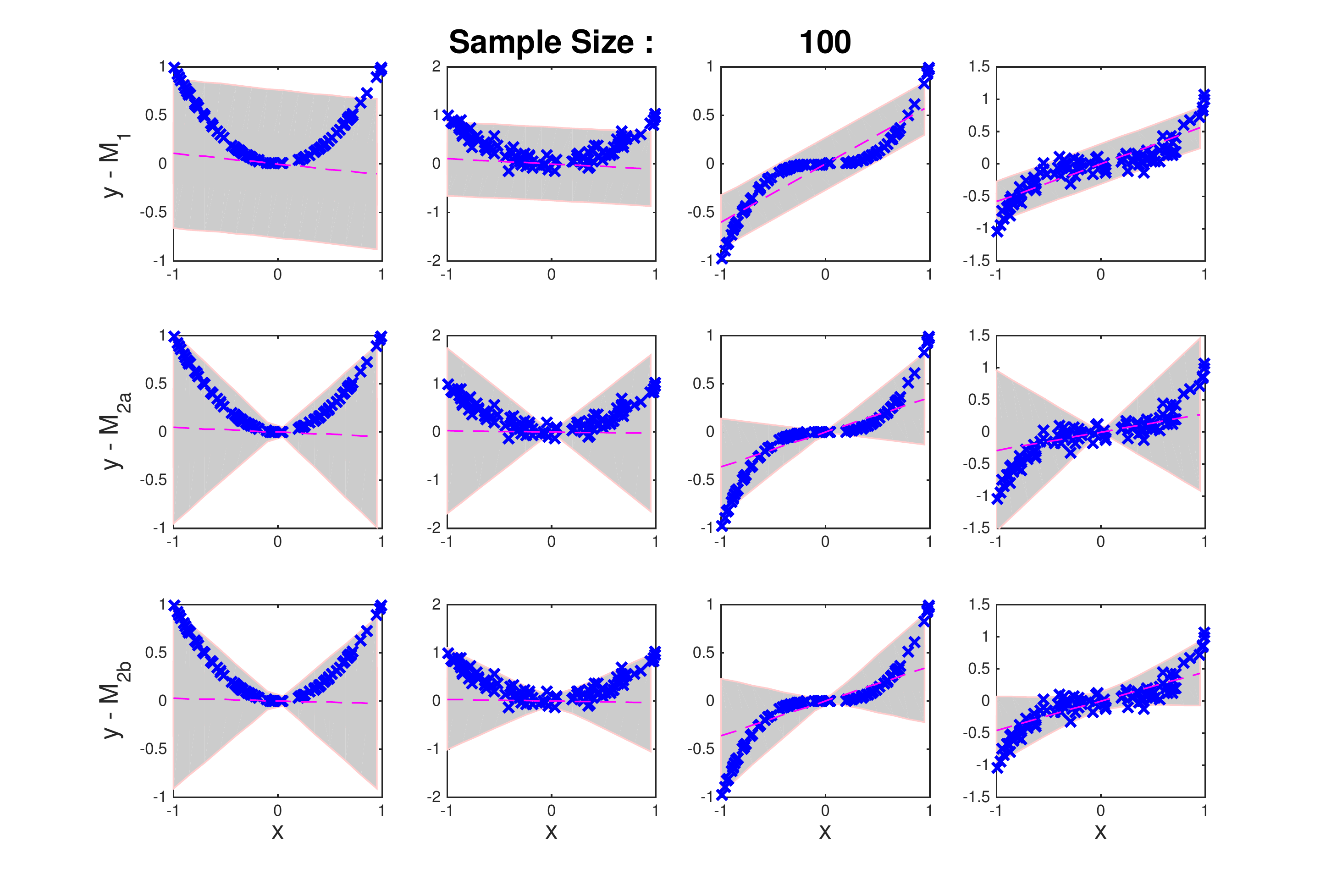}
	\caption{Posterior robust prediction for different models in different cases (sample size = 100). The purple dash lines denote the mean prediction and the grey area encloses 90\% of the total probability density for the predicted value. Blue crosses are the data points. The four columns correspond to those in Table \ref{tab:C3}.}
	\label{fig:C3_d3}
\end{figure}

Figure \ref{fig:C3_d3} shows one set of the prediction results (the results are not sensitive to the change of the number of sample size). We observe that for cases without additive noise, predictions from $\mM_{2a}$ and $\mM_{2b}$ are almost the same. Because $\mM_{2a}$ is a simpler model than $\mM_{2b}$ (less parameters), Bayesian model class selection prefers $\mM_{2a}$ and gives it a higher log-evidence value. For cases with additive noise, the uncertainty of prediction from $\mM_{2a}$ is significantly larger than $\mM_{2b}$. This is because $\theta$ is very sensitive to noise for $x$ values close to zero when a model does not have any additive error component. As a result, Bayesian model class selection prefers $\mM_{2b}$ for its ability to better fit the data on average. These results suggest that the HSM with additive error is preferred as the first test for uncertainty quantification of reduced order model. The flexibility of such a model class to handle two types of uncertainty simultaneously (additive error and embedded error in the model parameters) results in a higher chance of discovering the underlying uncertainty structure of a reduced order model. 

%\begin{figure}
%	\centering
%	%\input{figures/fig_Data}
%	\includegraphics[width=\textwidth]{figures/C3_d1.eps}
%	\caption{Robust prediction for different models in different cases (sample size = 20). The purple denotes the mean prediction and the grey area encloses 90\% of the total probability density for the predicted value. Blue crosses are the data points.}
%	\label{fig:C3_d1}
%\end{figure}
%
%\begin{figure}
%	\centering
%	%\input{figures/fig_Data}
%	\includegraphics[width=\textwidth]{figures/C3_d2.eps}
%	\caption{Robust prediction for different models in different cases (sample size = 50). The purple denotes the mean prediction and the grey area encloses 90\% of the total probability density for the predicted value. Blue crosses are the data points.}
%	\label{fig:C3_d2}
%\end{figure}

%\begin{figure}
%	\centering
%	%\input{figures/fig_Data}
%	\includegraphics[width=\textwidth]{figures/C3_d4.eps}
%	\caption{Robust prediction for different models in different cases (sample size = 200). The purple denotes the mean prediction and the grey area encloses 90\% of the total probability density for the predicted value. Blue crosses are the data points.}
%	\label{fig:C3_d4}
%\end{figure}

\section{Efficient Approximation of HSM}\label{sec:approx}

The posterior distribution $p(\vpsi|\mD)$ plays an important role in the HSM. Equation \ref{eq:HSM_PP} shows that the calculation of the likelihood $p(\mD|\vpsi)$ involves evaluations of multiple integrals, which correspond to the evidences for each data set $D_i$ conditional on the hyperparameters $\vpsi$. This leads to an extremely large computational cost for sampling from $p(\vpsi|\mD)$, as well as estimating the evidence of the model, $p(\mD)$. Current approaches include using conjugate pairs for analytical results \cite{Congdon:2010}, approximating the integrals with Laplace Asymptotic Approximation \cite{Wu+et_al:2015}, or using some advanced Markov Chain Monte Carlo techniques \cite{Nagel+Sudret:2015}. These methods are still restrictive for either studying many complex systems or the efficiency is not very scalable to handle extra data sets.

In this section, we present an efficient approximation method based on a special use of Importance Sampling. Most current research efforts focus on the HSM that adds only one extra level to the classical Bayesian model. Our method can be intuitively and efficiently extended to any complex HSMs with more levels of parameters. Moreover, we can lay out a standard procedure for fully analyzing any HSMs based on this method. 

\begin{figure}
	\centering
	\begin{subfigure}[t]{0.4\textwidth}
		\centering
		\includegraphics[width=\textwidth]{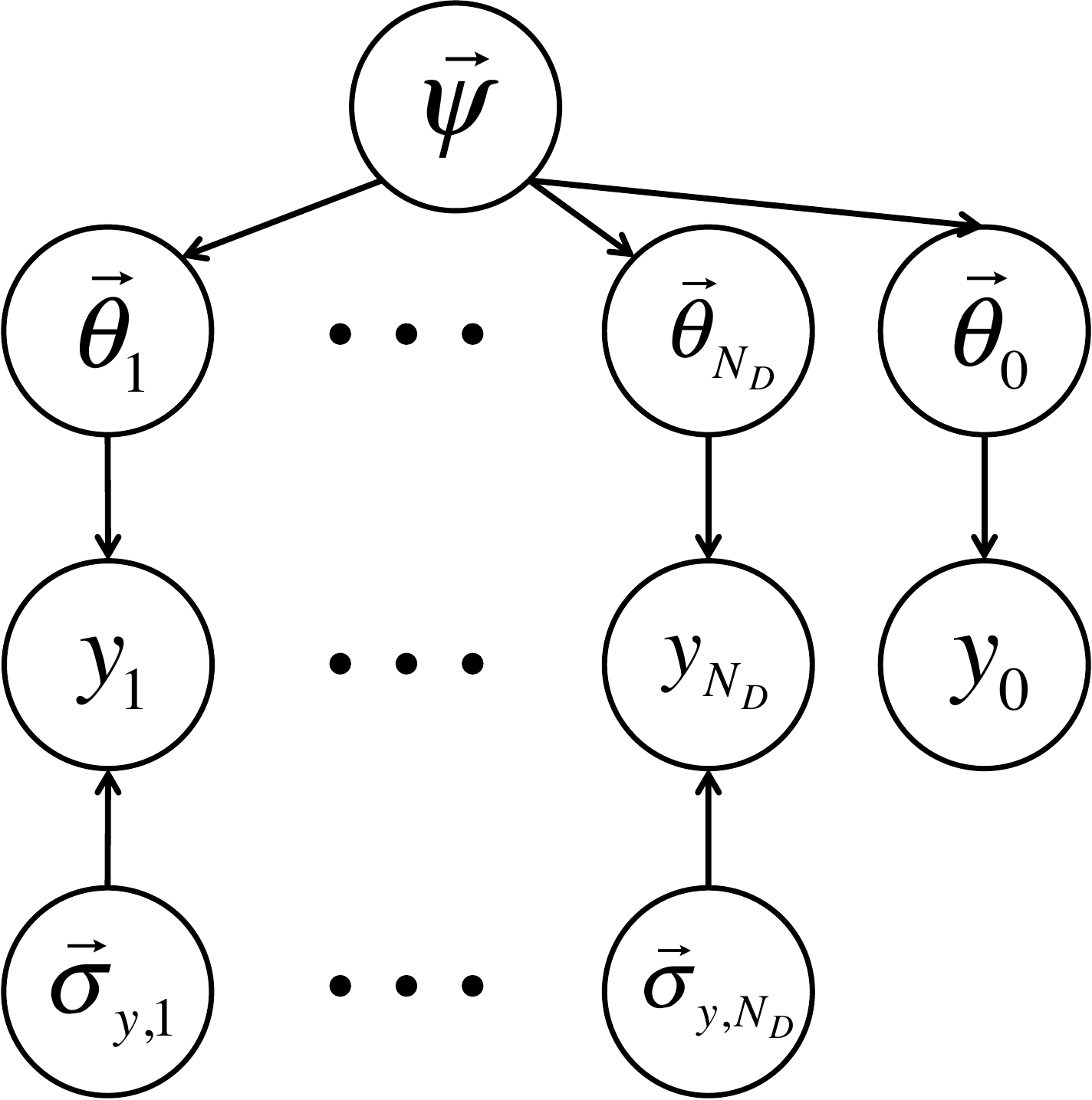}
		\caption{Independent additive error parameters}
		\label{fig:BN_HBa1}
	\end{subfigure}
	\qquad
	\begin{subfigure}[t]{0.4\textwidth}
		\centering
		\includegraphics[width=\textwidth]{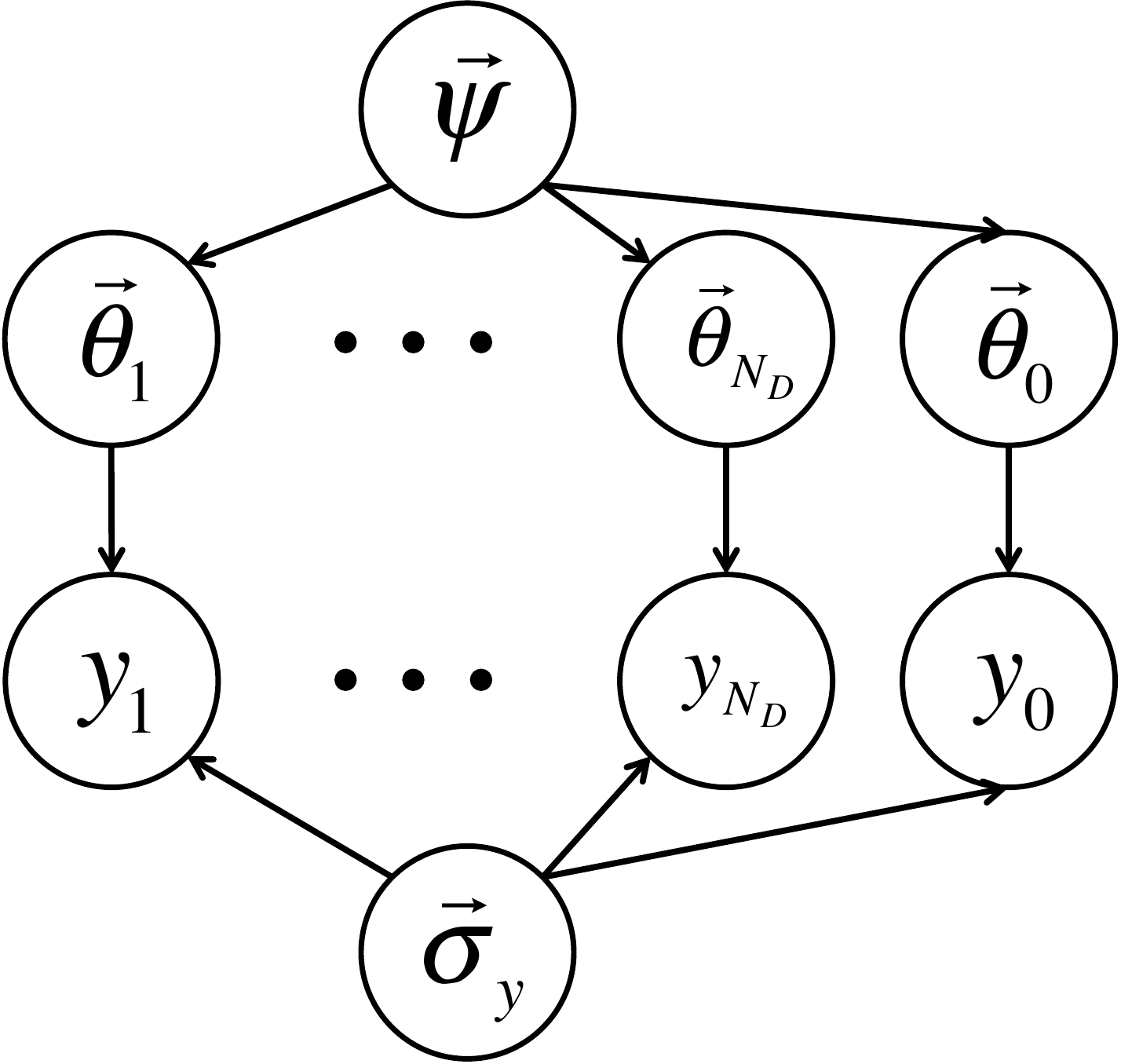}
		\caption{Common additive error parameters}
		\label{fig:BN_HBa2}
	\end{subfigure}
	\caption{Graphical representations of the two alternative HSMs.}
	\label{fig:BN_HBs2}
\end{figure}

For ease of illustration, we demonstrate our method based on the commonly used stochastic model shown in Equation \ref{eq:basicSM}. In practice, there may be more complicated HSM than the one shown in Figure \ref{fig:BN_HB}. Here, we consider two possible alternatives (see Figure \ref{fig:BN_HBs2}): (1) independent additive error parameters --- applicable to heterogeneous data with different likelihoods \cite{Wu_uncecomp:2015}, and (2) common additive error parameters --- applicable to data sets that are expected to have the same measurement errors \cite{Wu+et_al:2015}. 

We introduce our method by first applying it to the basic HSM shown in Figure \ref{fig:BN_HB}, which represents the cases that either $\vsigma_y$ is known or it is uncertain and included with the other uncertain parameters $\vtheta$. Then, we extend the idea to the two alternative HSMs shown in Figure \ref{fig:BN_HBs2}.

\subsection{Basic HSM}
This model class considers $\sigma_y$ to have a known value or to be treated as part of $\vtheta$. Hence, we do not infer the distribution of $\sigma_y$ from the data $\mD$ separately. Figure \ref{fig:BN_HB} shows the graphical representation for this case. In order to separate it from the two alternative HSMs in Figure \ref{fig:BN_HBs2}, we denote this model class as $\mM_{HS1}$. The posterior distribution, the likelihood and the evidence for $\vpsi$ are:
\begin{align}
p(\psi|\mD,\mM_{HS1}) &= \frac{p(\mD|\psi,\mM_{HS1})p(\psi|\mM_{HS1})}{p(\mD|\mM_{HS1})}\\
p(\mD|\psi,\mM_{HS1}) &= \prod_{i=1}^{N_D} p(D_i|\psi,\mM_{HS1}) \nonumber\\
&= \prod_{i=1}^{N_D} \int \! p(D_i|\theta_i,\mM_{HS1})p(\theta_i|\psi,\mM_{HS1}) \, \mathrm{d}\theta_i \label{eq:lik_psi}\\
p(\mD|\mM_{HS1}) &= \int \! p(\mD|\psi,\mM_{HS1})p(\psi|\mM_{HS1}) \, \mathrm{d}\psi
\end{align}

Sampling from $p(\psi|\mD,\mM_{HS1})$ requires repeated evaluations of the likelihood $p(\mD|\psi,\mM_{HS1})$ and thus evaluations of the integrals with different values of $\vpsi$. Our idea is to approximate the integrals by Importance Sampling with a special choice of the proposal PDF that can significantly reduce the total computational cost. In many cases, the likelihood $p(D_i|\theta_i,\mM_{HS1})$, which is related to the evaluation of $f(x,\vtheta)$, dominates the computational effort. 

Our method begins by performing Bayesian inference for each data set $D_i$ using the same likelihood $p(D_i|\theta_i,\mM_{HS1})$, but a different prior (choice of such prior is discussed later). To be more specific, we draw samples $\{\vtheta_i^{(j)} | j = 1,\dots,N_{s,i}\}$ from the posterior distribution $p(\vtheta_i | D_i,\mM_i)$, where $\mM_i$ denotes this specific stochastic model class:
\begin{equation}
\begin{aligned}
&p(\vtheta_i | D_i,\mM_i) = \frac{p(D_i|\vtheta_i,\mM_i)p(\vtheta|\mM_i)}{p(D_i|\mM_i)}\\
&\text{where} \,\,\, p(D_i|\vtheta_i,\mM_i) = p(D_i|\vtheta_i,\mM_{HS1})
\end{aligned}
\end{equation}
Then, we can approximate $p(D_i|\psi,\mM_{HS1})$ based on IS with proposal distribution $q_i(\vtheta_i) = p(\vtheta_i | D_i,\mM_i)$:
\begin{equation}
\begin{aligned}
p(D_i|\psi,\mM_{HS1}) &\approx \frac{1}{N_{s,i}} \sum_{j=1}^{N_{s,i}} \frac{p(D_i|\vtheta_i^{(j)},\mM_{HS1})p(\vtheta_i^{(j)}|\psi,\mM_{HS1})}{q_i(\vtheta_i^{(j)})}\\
&= \frac{p(D_i|M_i)}{N_{s,i}} \sum_{j=1}^{N_{s,i}} \frac{p(\vtheta_i^{(j)}|\psi,\mM_{HS1})}{p(\vtheta_i^{(j)}|\mM_i)}\\
&\text{where} \,\,\, \vtheta_i^{(j)} \sim p(\vtheta_i | D_i,\mM_i)
\end{aligned}
\end{equation}
As a result, we only need to perform classical Bayesian inference once for each data set $D_i$ (draw posterior samples and estimate the evidence $p(D_i|\mM_i)$). Then, the hierarchical analysis comes as a post-processing of the results. A very good tool for such a Bayesian inference is the Transitional Markov Chain Monte Carlo (TMCMC) method, where the evidence comes as a by-product of drawing posterior samples and the algorithm is inherently parallel \cite{Ching+Chen:2007}. The likelihood $p(\mD|\psi,\mM_{HS1})$ can then be estimated by:
\begin{equation}
p(\mD|\vpsi,\mM_{HS1}) \approx \prod_{i=1}^{N_D} \left( \frac{p(D_i|M_i)}{N_{s,i}} \sum_{j=1}^{N_{s,i}} \frac{p(\vtheta_i^{(j)}|\vpsi,\mM_{HS1})}{p(\vtheta_i^{(j)}|\mM_i)} \right), \,\,\, \text{where} \,\,\, \vtheta_i^{(j)} \sim p(\vtheta_i | D_i,\mM_i)
\end{equation}
In other words, the evidence ratio between $\mM_{HS1}$ with given hyperparameters and $\mM_i$ can be approximated by the mean of the prior ratio between the two models over the posterior $\vtheta_i$ samples of $\mM_i$.

This approximation suffers the same problem as IS, i.e., the variance of the estimate depends strongly on the closeness of the integrand and the proposal distribution. In our case, they are exactly the same if $p(\vtheta_i|\mM_i) = p(\vtheta_i|\vpsi,\mM_{HS1})$. Therefore, we should choose $p(\vtheta_i|\mM_i)$ to minimize its difference to $p(\vtheta_i|\vpsi,\mM_{HS1})$ for all $\vpsi$ weighted by the prior $p(\vpsi|\mM_{HS1})$. If we use the Kullback-Leibler divergence $D_{KL}(p||q)$ as a measure of difference between two distribution $p(x)$ and $q(x)$, this implies the choice:
\begin{equation}
\begin{aligned}
&p(\vtheta_i|\mM_i) = \arg\!\min \int \! D_{KL}(p(\vtheta_i|\vpsi,\mM_{HS1})||p(\vtheta_i|\mM_i)) p(\vpsi|\mM_{HS1}) \, \mathrm{d}\vpsi\\
&\text{where}~~ D_{KL}(p(x)||q(x)) = \int \! p(x)\log\frac{p(x)}{q(x)} \, \mathrm{d}x
\end{aligned}
\end{equation}

A large number of posterior samples from $p(\vtheta_i|D_i,\mM_i)$ can also reduce the variance of the IS estimate. If this is computationally feasible, an alternative for $p(\theta_i|D_i,\mM_i)$ would be a uniform distribution that covers the significant regions of $p(\vtheta_i|\mM_{HS1}) = \int \! p(\vtheta_i|\vpsi,\mM_{HS1})p(\vpsi|\mM_{HS1}) \, \mathrm{d}\vpsi$. We note that the evidence will be insignificantly small for $\vpsi$ values that lead to the prior $p(\vtheta_i|\vpsi,\mM_{HS1})$ having a mode well away from the maximum of the likelihood $p(D_i|\vtheta_i,\mM_{HS1})$. Hence, accuracy is not important in this case. In the opposite case, IS is a good approximation. The inaccuracy problem is important only for the middle case, i.e., the prior mode is not too far and not too close to the maximum of the likelihood. The range of $\vpsi$ leading to this important case depends on the posterior sample size $N_{s,i}$ for each data set $D_i$.
 
We note that this approximation method provides information about Bayesian inference for each individual data set first. This information is useful to compare with the HSM analysis to give further insight about the system of interest. Furthermore, our method is very efficient for introducing extra data sets because we do not need to rerun the whole problem. Instead, the overhead is only the classical Bayesian inferences for the extra data sets and a post-processing step that is not computationally intensive. 

\subsection{HSM with independent additive error parameters}
This model class considers $\sigma_y$ for the predictions of each data set to be independent of each other. Figure \ref{fig:BN_HBa1} shows the graphical representation for this case. We denote this model class as $\mM_{HS2}$. The posterior distribution, the likelihood and the evidence for $\vpsi$ are:
\begin{align}
p(\vpsi|\mD,\mM_{HS2}) &= \int \! p(\vpsi,\vsigma_{y,1},\dots,\vsigma_{y,N_D}|\mD,\mM_{HS2}) \, \mathrm{d}\vsigma_y \dots \mathrm{d}\vsigma_{y,N_D} \label{eq:HS2_P}\\
p(\vpsi,\vsigma_{y,1},\dots,\vsigma_{y,N_D}|\mD,\mM_{HS2}) &= \frac{p(\mD|\vpsi,\vsigma_{y,1},\dots,\vsigma_{y,N_D},\mM_{HS2})p(\psi,\vsigma_{y,1},\dots,\vsigma_{y,N_D}|\mM_{HS2})}{p(\mD|\mM_{HS2})} \label{eq:HS2_PS}\\
p(\mD|\vpsi,\vsigma_{y,1},\dots,\vsigma_{y,N_D},\mM_{HS2}) &= \prod_{i=1}^{N_D} \int \! p(D_i|\vtheta_i,\vsigma_{y,i},\mM_{HS2})p(\vtheta_i|\vpsi,\mM_{HS2}) \, \mathrm{d}\vtheta_i
\end{align}
\begin{equation}
\begin{aligned}
&p(\mD|\mM_{HS2}) \\
= &\int \! p(\mD|\vpsi,\vsigma_{y,1},\dots,\vsigma_{y,N_D},\mM_{HS2})p(\vpsi,\vsigma_{y,1},\dots,\vsigma_{y,N_D}|\mM_{HS2}) \, \mathrm{d}\vpsi \mathrm{d}\vsigma_{y,1} \dots \mathrm{d}\vsigma_{y,N_D}
\end{aligned}
\end{equation}

The idea for $\mM_{HS1}$ does not apply to this model class directly because both the likelihood $p(D_i|\vtheta_i,\vsigma_{y,i},\mM_{HS2})$ and the prior $p(\vtheta_i|\vpsi,\mM_{HS2})$ will be affected by the sampling of $\vpsi, \vsigma_{y,1}, \dots, \vsigma_{y,N_D}$. However, if prediction is the ultimate goal, Figure \ref{fig:BN_HBa1} implies that $\vsigma_{y,1},\dots,\vsigma_{y,N_D}$ are not relevant as long as the posterior distribution of $\vpsi$ is known. Then, we can again use IS to estimate $p(\vpsi|\mD,\mM_{HS2})$ directly.

We note that this is a special case of $\mM_{HS1}$ with $p(\vtheta_i,\vsigma_{y,i}|\vpsi) = p(\vtheta_i|\vpsi)p(\vsigma_{y,i})$, i.e., $\vsigma_{y,i}$ is independent of $\vpsi$. Similar to $\mM_{HS1}$, we first draw samples $\{ (\vtheta_i^{(j)},\vsigma_{y,i}^{(j)}) | j = 1,\dots,N_{s,i}\}$ from the posterior distribution $p(\vtheta_i,\vsigma_{y,i} | D_i,\mM_i)$:
\begin{equation}
\begin{aligned}
&p(\vtheta_i,\vsigma_{y,i}| D_i,\mM_i) = \frac{p(D_i|\vtheta_i,\vsigma_{y,i},\mM_i)p(\vtheta_i|\mM_i)p(\vsigma_{y,i}|\mM_i)}{p(D_i|\mM_i)}\\
&\text{where} \,\,\, p(D_i|\vtheta_i,\vsigma_{y,i},\mM_i) = p(D_i|\vtheta_i,\vsigma_{y,i},\mM_{HS2})
\end{aligned}
\end{equation}
%Then, combining Equation \ref{eq:HS2_P} and \ref{eq:HS2_PS} and using the fact that: 
%\begin{equation}
%p(\vpsi,\vsigma_{y,1},\dots,\vsigma_{y,N_D}|\mM_{HS2}) = p(\vpsi|\mM_{HS2})\prod_{i=1}^{N_D} p(\vsigma_{y,i}|\mM_{HS2})
%\end{equation}
%we can rewrite $p(\vpsi|\mD,\mM_{HS2})$ as:
%\begin{equation}
%\begin{aligned}
%p(\vpsi|\mD,\mM_{HS2}) &= \frac{\left(\prod_{i=1}^{N_D} \int \! p(D_i|\vtheta_i,\vsigma_{y,i},\mM_{HS2})p(\vtheta_i|\vpsi,\mM_{HS2})p(\vsigma_{y,i}|\mM_{HS2}) \, \mathrm{d}\vtheta_i \mathrm{d}\vsigma_{y,i} \right) p(\vpsi|\mM_{HS2})}{p(\mD|\mM_{HS2})}\\
%&= \frac{\left(\prod_{i=1}^{N_D} p(D_i|\vpsi,\mM_{HS2}) \right) p(\vpsi|\mM_{HS2})}{p(\mD|\mM_{HS2})}
%\end{aligned}
%\end{equation}
%Hence, we can approximate $p(D_i|\vpsi,\mM_{HS2})$ based on IS with proposal distribution $q_i(\vtheta_i,\vsigma_{y,i}) = p(\vtheta_i,\vsigma_{y,i} | D_i,\mM_i)$:
%\begin{equation}
%\begin{aligned}
%p(D_i|\vpsi,\mM_{HS2}) &= \int \! p(D_i|\vtheta_i,\vsigma_{y,i},\mM_{HS2})p(\vtheta_i|\vpsi,\mM_{HS2}) \, \mathrm{d}\vtheta_i\\
%&\approx \frac{1}{N_{s,i}} \sum_{j=1}^{N_{s,i}} \frac{p(D_i|\vtheta_i^{(j)},\vsigma_{y,i}^{(j)},\mM_{HS2})p(\vtheta_i^{(j)}|\vpsi,\mM_{HS2})}{q_i(\vtheta_i^{(j)},\vsigma_{y,i}^{(j)})}\\
%&= \frac{p(D_i|M_i)}{N_{s,i}} \sum_{j=1}^{N_{s,i}} \frac{p(\vtheta_i^{(j)}|\vpsi,\mM_{HS2})}{p(\vtheta_i^{(j)}|\mM_i)} \frac{p(\vsigma_{y,i}^{(j)}|\mM_{HS2})}{p(\vsigma_{y,i}^{(j)}|\mM_i)}\\
%&\text{where} \,\,\, (\vtheta_i^{(j)},\vsigma_{y,i}^{(j)}) \sim p(\vtheta_i,\vsigma_{y,i} | D_i,\mM_i)
%\end{aligned}
%\end{equation}
Following the procedure for $\mM_{HS1}$, we can derive that:
\begin{equation}
\begin{aligned}
p(D_i|\vpsi,\mM_{HS2}) &\approx \frac{p(D_i|M_i)}{N_{s,i}} \sum_{j=1}^{N_{s,i}} \frac{p(\vtheta_i^{(j)}|\vpsi,\mM_{HS2})}{p(\vtheta_i^{(j)}|\mM_i)} \frac{p(\vsigma_{y,i}^{(j)}|\mM_{HS2})}{p(\vsigma_{y,i}^{(j)}|\mM_i)}\\
&\text{where} \,\,\, (\vtheta_i^{(j)},\vsigma_{y,i}^{(j)}) \sim p(\vtheta_i,\vsigma_{y,i} | D_i,\mM_i)
\end{aligned}
\end{equation}
Again, the HSM analysis comes as a post-processing step and $p(\mD|\vpsi,\mM_{HS2}) = \prod_{i=1}^{N_D} p(D_i|\vpsi,\mM_{HS2})$ can then be estimated using the posterior samples $(\vtheta_i^{(j)},\vsigma_{y,i}^{(j)})$:
\begin{equation}\label{eq:HS2_D}
p(D_i|\vpsi,\mM_{HS2}) \approx \prod_{i=1}^{N_D} \left( \frac{p(D_i|M_i)}{N_{s,i}} \sum_{j=1}^{N_{s,i}} \frac{p(\vtheta_i^{(j)}|\vpsi,\mM_{HS2})}{p(\vtheta_i^{(j)}|\mM_i)} \frac{p(\vsigma_{y,i}^{(j)}|,\mM_{HS2})}{p(\vsigma_{y,i}^{(j)}|\mM_i)} \right)
\end{equation}
As a result, we can directly obtain posterior samples from $p(\vpsi|\mD,\mM_{HS2})$ using MCS or MCMC methods combined with Equation \ref{eq:HS2_D}. We note that the estimation can be further simplified and more accurate when choosing $p(\sigma_{y,i}|\mM_i) = p(\sigma_{y,i}|\mM_{HS2})$ for all $i$.

\subsection{HSM with common additive error parameters}
This model class considers a single value of $\vsigma_y$ to be shared by all predictions. Figure \ref{fig:BN_HBa2} shows the graphical representation for this case. We denote this model class as $\mM_{HS3}$. In this model class, $\vsigma_y$ affects future prediction as well. Hence, the approach for $\mM_{HS2}$ will not work. We adopt the idea from EIM to solve the problem. The basic concept of our approach is to approximate the likelihood function $p(D_i|\vtheta_i,\vsigma_y,\mM_{HS3})$ as a linear sum of multiple copies of the same term, but with fixed $\vsigma_y$ values. Then, we can perform the classical Bayesian analyses for those fixed $\vsigma_y$ cases and use the same IS approach as in $\mM_{HS1}$ and $\mM_{HS2}$ to approximate the HSM analysis for $\mM_{HS3}$. First, we look at how to use the EIM idea to estimate the posterior distributions in $\mM_{HS3}$. Then, we discuss how to train the linear approximations for the likelihood $p(D_i|\vtheta_i,\vsigma_y,\mM_{HS3})$.

We denote $g_l(\vtheta_i) = p(D_i|\vtheta_i,\vsigma_{y,l},\mM_{HS3})$ as a basis function for some fixed value of $\vsigma_{y,l}$ for $l=1,\dots,L$. EIM assumes the following approximation for the fixed number of bases $L$ and some coefficients $\alpha_l(\vsigma_y)$:
\begin{equation}\label{eq:EIM0}
p(D_i|\vtheta_i,\vsigma_y,\mM_{HS3}) \approx \sum_{l=1}^L \alpha_l(\vsigma_y) g_l(\vtheta_i)
\end{equation}
Then, we can approximate the likelihood $p(D_i|\vpsi,\vsigma_y,\mM_{HS3}) = \int \! p(D_i|\vtheta_i,\vsigma_y,\mM_{HS3})p(\vtheta_i|\vpsi,\mM_{HS3}) \, \mathrm{d}\vtheta_i$ using Equation \ref{eq:EIM0}:
\begin{equation}\label{eq:EIM1}
p(D_i|\vpsi,\vsigma_y,\mM_{HS3}) \approx \sum_{l=1}^L \alpha_l(\vsigma_y) \int \! p(D_i|\vtheta_i,\vsigma_{y,l},\mM_{HS3})p(\vtheta_i|\vpsi,\mM_{HS3}) \, \mathrm{d}\vtheta_i
\end{equation}
Since $\vsigma_{y,l}$ is fixed, we can apply the same IS approach as in $\mM_{HS2}$ and $\mM_{HS1}$ to the integrals in Equation \ref{eq:EIM1}. We draw samples $\{\vtheta_i^{(j)} | j = 1,\dots,N_{s,i}\}$ from $p(\vtheta_i | D_i,\vsigma_{y,l},\mM_i)$:
\begin{equation}
\begin{aligned}
&p(\vtheta_i | D_i,\vsigma_{y,l},\mM_i) = \frac{p(D_i|\vtheta_i,\vsigma_{y,l},\mM_i)p(\vtheta_i|\mM_i)}{p(D_i|\vsigma_{y,l},\mM_i)}\\
&\text{where} \,\,\, p(D_i|\vtheta_i,\vsigma_{y,l},\mM_{HS3}) = p(D_i|\vtheta_i,\vsigma_{y,l},\mM_i)
\end{aligned}
\end{equation}
Then, we use the proposal $q_{l,i}(\vtheta_i) = p(\vtheta_i | D_i,\vsigma_{y,l},\mM_i)$:
\begin{equation}
\begin{aligned}
p(D_i|\vpsi,\vsigma_{y,l},\mM_{HS3}) &\approx \frac{1}{N_{s,i,l}} \sum_{j=1}^{N_{s,i,l}} \frac{p(D_i|\vtheta_{i,l}^{(j)},\vsigma_{y,l},\mM_{HS3})p(\vtheta_{i,l}^{(j)}|\vpsi,\mM_{HS3})}{q_{l,i}(\vtheta_{i,l}^{(j)})}\\
&= \frac{p(D_i|\vsigma_{y,l},M_i)}{N_{s,i,l}} \sum_{j=1}^{N_{s,i,l}} \frac{p(\vtheta_{i,l}^{(j)}|\vpsi,\mM_{HS3})}{p(\vtheta_{i,l}^{(j)}|\mM_i)}\\
&\text{where} \,\,\, \vtheta_{i,l}^{(j)} \sim p(\vtheta_i | D_i,\vsigma_{y,l},\mM_i)
\end{aligned}
\end{equation}
As a result, the likelihood $p(\mD|\vpsi,\vsigma_y,\mM_{HS3})$ can, then, be estimated by:
\begin{equation}\label{eq:HS3_EIM}
\begin{aligned}
&p(\mD|\vpsi,\vsigma_y,\mM_{HS3}) \approx \prod_{i=1}^{N_D} \left( \sum_{l=1}^L \alpha_l(\vsigma_y) \frac{p(D_i|\vsigma_{y,l},M_i)}{N_{s,i,l}} \sum_{j=1}^{N_{s,i,l}} \frac{p(\vtheta_{i,l}^{(j)}|\vpsi,\mM_{HS3})}{p(\vtheta_{i,l}^{(j)}|\mM_i)} \right)\\
&\text{where} \,\,\, \vtheta_{i,l}^{(j)} \sim p(\vtheta_i | D_i,\vsigma_{y,l},\mM_i)
\end{aligned}
\end{equation}
With this analytical expression for the approximation, the remaining problem is how to pick $\vsigma_{y,l}$ and how to find $\alpha_l(\vsigma_y)$ when estimating the posterior distribution $p(\vpsi,\vsigma_y|\mD,\mM_{HS3})$.

\subsubsection{Training basis functions}\label{sec:EIM_tr}
Hesthaven et al. \cite{Hesthaven:2014kx} suggest using an adaptive greedy algorithm to select the set of basis functions $g_l(\theta_i)$. Based on this idea, we develop an algorithm that simultaneously selects basis functions and collects posterior samples used in Equation \ref{eq:HS3_EIM}. The greedy algorithm reduces the maximum absolute value of the error term $e_l(\vtheta_i,\vsigma_{y,l})$ to some specified threshold $\tilde{\epsilon}_{lim}$ over training sets of $\vtheta_i$ and $\vsigma_y$, denote as $\Theta_i$ and $\Sigma_y$, respectively. The efficiency and accuracy of the algorithm on error estimation is a tradeoff that depends on the size of the training sets. The error term corresponding to $L$ basis functions is defined as:
\begin{equation}
e_L(\vtheta_i,\vsigma_y) = |p(D_i|\vtheta_i,\vsigma_y,\mM_{HS3}) - \sum_{l=1}^L \alpha_l(\vsigma_y) g_l(\vtheta_i)|
\end{equation}

Starting with some initial training sets $\Theta_i^{initial}$ and $\Sigma_y^{initial}$ and an initial set of basis functions $G^{initial}$, a new basis function with a corresponding $\vsigma_y$ value is chosen from $\Sigma_y$ to maximize the error term $e_L$. Also, we record the $\vtheta_i$ values that maximize $e_L$ for the chosen $\vsigma_y$. These values will be used in the ``online" estimation stage discussed later. Then, a set of posterior samples $\{\vtheta_i\}_L$ is drawn using any MCMC methods with the chosen $\vsigma_y$, and the samples are added to the current training sets $\Theta_i$. This process is repeated until the maximum error given $\Theta_i$ is below the threshold $e_{lim}$. In the end, we obtain a set of bases defined by $(\vtheta_{i,l}, \vsigma_{y,l})$, $l=1,\dots,L$, the posterior samples $\{\vtheta_i\}_l$ and the evidence values $p(D_i|\vsigma_{y,l},\mM_{HS3})$ for estimation of the approximate hierarchical Bayesian inference. Algorithm \ref{alg:train} summarizes this new adaptive training procedure.

\begin{algorithm}[H]
	\caption{: Adaptive greedy algorithm}\label{alg:train}
	\begin{algorithmic}
		\STATE Obtain initial $\vsigma_{y,1},\dots,\vsigma_{y,L_0}$ for basis functions $g_1,\dots,g_{L_0}$ from $G^{initial}$
		\STATE Obtain initial $\vtheta_{i,1},\dots,\vtheta_{i,L_0}$ for the corresponding basis functions from $G^{initial}$
		\STATE Initialize counter $l \leftarrow L_0$
		\STATE Initialize training sets $\Theta_i = \Theta_i^{initial}$ and $\Sigma_y = \Sigma_y^{initial}$
		\STATE $\tilde{\epsilon}_{L_0} = \max_{\vsigma_y \in \Sigma_y} \max_{\vtheta_i \in \Theta_i} e_{L_0}(\vtheta_i,\vsigma_y)$
		\WHILE{$\tilde{\epsilon}_l > \tilde{\epsilon}_{lim}$}
		\STATE $l \leftarrow l + 1$
		\STATE $\vsigma_{y,l} = \mathrm{argmax}_{\vsigma_y \in \Sigma_y} \left\{ \max_{\vtheta_i \in \Theta_i} e_{l-1}(\vtheta_i,\vsigma_y) \right\}$
		\STATE $\vtheta_{i,l} = \mathrm{argmax}_{\vtheta_i \in \Theta_i} e_{l-1}(\vtheta_i,\vsigma_{y,l})$
		\STATE $g_{l} (\vtheta_i) = p(D_i|\vtheta_i,\vsigma_{y,l},\mM_{HS3})$
		\STATE Obtain posterior samples $\{\vtheta_i\}_l$ from distribution $p(\vtheta_i|D_i,\vsigma_{y,l},\mM_i)$ using a MCMC method and calculate/record the evidence value $p(D_i|\vsigma_{y,l},\mM_i)$
		\STATE $\Theta_i \leftarrow \Theta_i \cup \{\vtheta_i\}_l$
		\STATE $\tilde{\epsilon}_{l} = \max_{\vsigma_y \in \Sigma_y} \max_{\vtheta_i \in \Theta_i} e_{l}(\vtheta_i,\vsigma_y)$
		\ENDWHILE
	\end{algorithmic}
\end{algorithm}

By enriching the training set $\Theta_i$, the error estimate $e_L(\vtheta_i,\vsigma_y)$ becomes more accurate and thus improves the EIM approximation. In many cases, most of the computational time for calculating $p(D_i|\vtheta_i,\vsigma_y,\mM_{HS3})$ is spent on evaluating the function $f(x,\vtheta_i)$. Because this evaluation is already done during MCMC sampling, the overhead of using a larger training set $\Theta_i$ is relatively small. However, it is important to note that the improvement of the error estimate may saturate. This occurs when the different sets of posterior samples for different $\vsigma_y$ values all have the same high probability regions. In this case, there will be many redundant samples in the training set clustering around the peaks of $p(D_i|\vtheta_i,\vsigma_y,\mM_{HS3})$. To further improve the efficiency of this training algorithm, we can control the expansion of $\Theta_i$ such that we only add samples that are significantly different from the samples in the current $\Theta_i$. A simple implementation is to monitor the spread of the chosen $\sigma_y$ because similar $\vsigma_y$ values represents a similar likelihood function value $p(D_i|\vtheta_i,\vsigma_y,\mM_{HS3})$ and thus similar posterior samples are expected. Using the same argument, we suggest constructing the initial sets $\Theta_i^{initial}$ and $G^{initial}$ based on extreme values of $\vsigma_y$ (e.g., maximum and minimum values of $\vsigma_y$ in 1D case). Algorithm \ref{alg:init} constructs the initial sets based on this suggestion.

\begin{algorithm}[H]
	\caption{: Constructing initial sets $\Theta_i^{initial}, \Sigma_y^{initial}, G^{initial}$}\label{alg:init}
	\begin{algorithmic}
		\STATE Pick a fine grid for $\Sigma_y^{initial}$
		\STATE Initialize $\vsigma_{y,1},\dots,\vsigma_{y,K}$ as a sequence of extreme $\vsigma_y$ values ordered in ascending ``peakness" of $p(D_i|\vtheta_i,\vsigma_y,\mM_{HS3})$, i.e., more probability content concentrated in a small region of $\vtheta_i$ values (only qualitatively, need not to be accurate). This is the initial set of $\vsigma_y$ values that define the initial basis functions $g_1(\vtheta_i),\dots,g_K(\vtheta_i)$ in $G^{initial}$
		\STATE Initialize $\Theta_i^{Initial} \leftarrow $ empty set or a set of sparse grid points
		\FOR{$k = 1$ \TO $K$}
		\IF{k = 1}
		\STATE $\vtheta_{i,k} = \mathrm{argmax}_{\vtheta_i \in \Theta_i} p(D_i|\vtheta_i,\vsigma_{y,k},\mM_{HS3})$
		\ELSE
		\STATE $\vtheta_{i,k} = \mathrm{argmax}_{\vtheta_i \in \Theta_i} e_{l-1}(\vtheta_i,\vsigma_{y,k})$
		\ENDIF
		\STATE Record $\vtheta_{i,k}$ in $G^{initial}$ corresponding to basis function $g_k(\vtheta_i)$
		\STATE Obtain posterior samples $\{\vtheta_i\}_k$ from distribution $p(\vtheta_i|D_i,\vsigma_{y,k},\mM_i)$ using a MCMC method and record the evidence value $p(D_i|\vsigma_{y,k},\mM_i)$
		\STATE $\Theta_i^{initial} \leftarrow \Theta_i^{initial} \cup \{\vtheta_i\}_k$
		\ENDFOR
	\end{algorithmic}
\end{algorithm}

\subsubsection{Online estimation}
Once we obtain the set of bases $(\vtheta_{i,l}, \vsigma_{y,l})$ that defines the basis functions $g_l(\vtheta_i)$, where $l = 1,\dots,L$ and $L$ is the total number of bases, we are ready to estimate $p(D_i|\vpsi^{(s)},\vsigma_y^{(s)},\mM_{HS3})$ with Equation \ref{eq:HS3_EIM} for any given hyperparameter sample $(\vpsi^{(s)},\vsigma_y^{(s)})$. First, we construct a matrix of the basis functions $g_{nl} = p(D_i|\vtheta_{i,n},\vsigma_{y,l},\mM_{HS3})$ for $n = 1,\dots,L$ and $l = 1,\dots,L$. Then, we calculate $P_n = p(D_i|\vtheta_{i,n},\vsigma_y^{(s)},\mM_{HS3})$, which is a vector of the actual values of the likelihood function. EIM constraints the linear approximation to be exact at the $\vtheta_i$ bases. Hence, we find the vector of all $\alpha_l(\vsigma_y^{(s)})$ by solving the linear equations:
\begin{equation}\label{eq:EIM_inv}
\begin{aligned}
&P_n = \sum_{l=1}^L \alpha_l (\vsigma_y^{(s)}) g_{nl}, \,\, 1 \leq n \leq L\\
\text{or} \,\,\,\, &\text{in matrix form :} \,\,\, [g_{nl}]\{\alpha_l\} = \{P_n\}
\end{aligned}
\end{equation}
Once $\alpha_l$ is solved, we can include the recorded posterior samples $\{\vtheta_i\}_l$ and evidence values $p(D_i|\vsigma_{y,l},M_i)$ corresponding to basis $g_l(\vtheta_i)$ to perform fast estimation of $p(D_i|\vpsi,\vsigma_y,\mM_{HS3})$.

\subsubsection{Numerical issue}
In Equation \ref{eq:EIM_inv}, note that the vector $\{P_n\}$ and each column of the matrix $[g_{nl}]$ represents the likelihood value $p(D_i|\vtheta_{i,n},\vsigma_y,\mM_{HS3})$ for different values of $\vsigma_y$. The value of the Gaussian likelihood is very sensitive to $\vsigma_y$ and thus the values of the columns of $[g_{nl}]$ and $\{P_n\}$ may be of different orders of magnitude. As a result, the matrix inversion in Equation \ref{eq:EIM_inv} often faces numerical issues because the matrix $[g_{nl}]$ is ill-conditioned. It is important to scale the matrix before solving the inversion problem. First, we rewrite the expression in Equation \ref{eq:EIM_inv} as:
\begin{equation}
\begin{aligned}
&\sum_{l=1}^L \left( c^\alpha_l \hat{\alpha}_l \right) \cdot c^g_l \{\hat{g}_l\} = c^P \{\hat{P}_n\}\\
&\text{where}~~~ c^P \{\hat{P}_n\} = \{P_n\}, [g_{nl}] = \left[ c^g_1 \{\hat{g}_1\} \cdots c^g_L \{\hat{g}_L\}\right] \\
&~~\text{and}~~~ \{\alpha_l\} = \{ c^{\alpha}_1 \hat{\alpha}_1 \cdots c^{\alpha}_L \hat{\alpha}_L \}^T
\end{aligned}
\end{equation}
If we choose $c^\alpha_l = c^P/c^g_l$ for all $l = 1,\dots,L$, the inversion problem becomes:
\begin{equation}\label{eq:EIM_inv2}
[\hat{g}_l] \{\hat{\alpha}_l\} = \{\hat{P}_n\}
\end{equation} 
We pick $c^P$ and $c^g_l$ to be the maximum value of the corresponding column vector. Then, the numerical issue due to scaling may not occur in Equation \ref{eq:EIM_inv2}. We can first solve the inversion problem in Equation \ref{eq:EIM_inv2} and recover the coefficients $\alpha_l$ by $\{\alpha_l\} = \{ c^{\alpha}_1 \hat{\alpha}_1 \cdots c^{\alpha}_L \hat{\alpha}_L \}^T$ where $c^\alpha_l = c^P/c^g_l$. Also, it may be useful to store the values in log-scale during the entire computation process.

\section{Illustrative Examples}\label{sec:ex}
We test our method using two examples: molecular dynamics simulation of Krypton and pharmacokinetic/pharmacodynamic (PKPD) modeling of a cancer drug. Both examples are taken from previous studies. Their results are used to validate the efficiency and accuracy of our method.

\subsection{Molecular Dynamics: Krypton}
Wu et al. \cite{Wu+et_al:2015} proposed using the HSM (called the HBM in the paper) to calibrate the Lennard-Jones (LJ) potential ($\epsilon_{LJ}$ and $\sigma_{LJ}$) of Krypton based on viscosity data $\mD$ reported from nine different laboratories. In that paper, the authors approximated the integrals in Equation \ref{eq:lik_psi} by Laplace Asymptotic Approximation (LAA). Since the joint posterior distribution of $\epsilon_{LJ}$ and $\sigma_{LJ}$ for each data set is close to a Gaussian distribution, LAA is a good approximation for this particular case. We re-visit the problem with the same data and the same model class setup as described in \cite{Wu+et_al:2015}. Instead of the LAA approximation, we use our proposed method to estimate the joint posterior distribution of $\epsilon_{LJ}$ and $\sigma_{LJ}$ and the posterior robust prediction. 

\begin{figure}[h]
	\centering
	\begin{subfigure}[h]{0.1\textwidth}
		\centering
		\includegraphics[width=\textwidth]{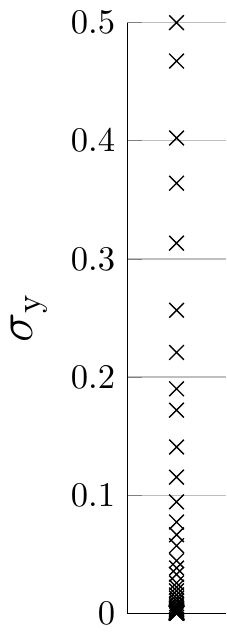}
	\end{subfigure}
	\begin{subfigure}[h]{0.4\textwidth}
		\centering
		\includegraphics[width=0.9\textwidth]{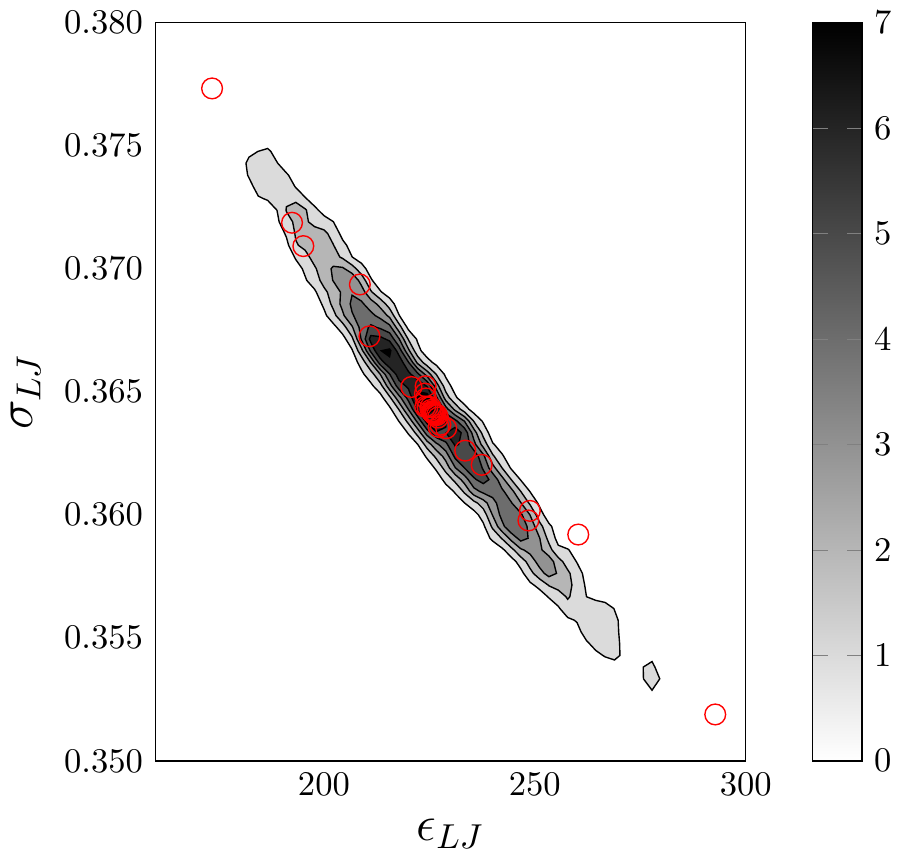}
	\end{subfigure}
 	\begin{subfigure}[h]{0.45\textwidth}
		\centering
		\includegraphics[width=0.75\textwidth]{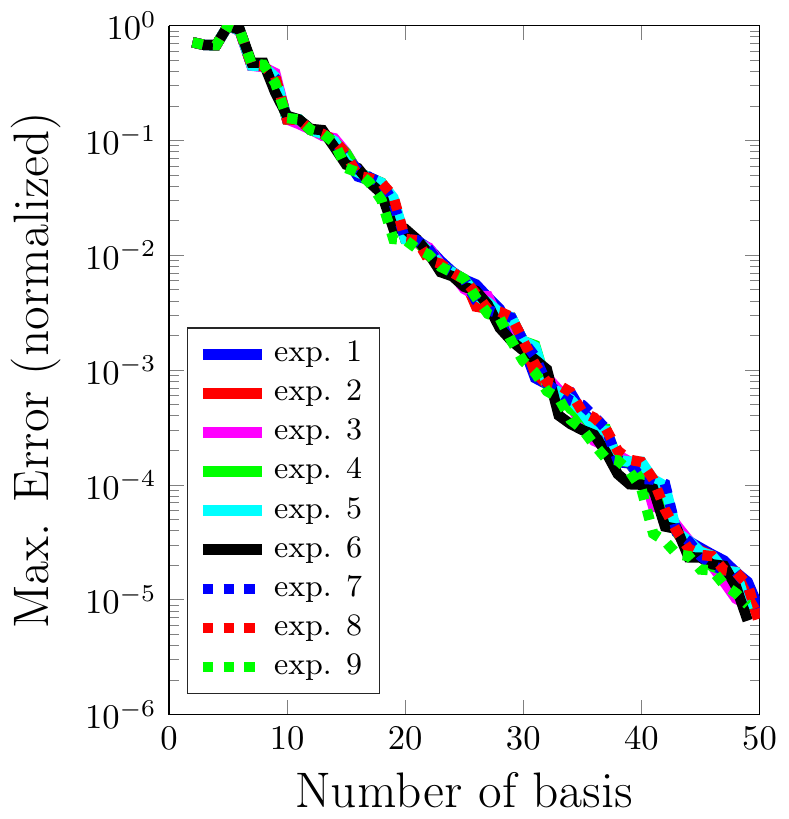}
	\end{subfigure}
	\caption{Results of the EIM approximation for the likelihood function. Left 2 plots: chosen bases of $\sigma_y$ (black cross), $\epsilon_{LJ}$ and $\sigma_{LJ}$ (red circles) for experiment 6. The gray scale contour shows the actual likelihood values for this experiment. Right plot: maximum error of the EIM estimate for each experiment as a function of the number of basis.  The error is normalized by the maximum value of the actual function.}
	\label{fig:Kr_EIM}
\end{figure}

Following the assumptions in \cite{Wu+et_al:2015}, we use the HSM with common additive error. First, we build the EIM basis functions based on Section \ref{sec:EIM_tr}. We begin with a training set of $\epsilon_{LJ}$ and $\sigma_{LJ}$ from a coarse grid of 256 points within the boundaries $100 \leq \epsilon_{LJ} \leq 400$ and $0.2 \leq \sigma_{LJ} \leq 0.5$. A fine grid of 172 points for $\sigma_y$ is used between 0.0001 and 0.5, uniformly distributed in log scale. We obtain a total of 50 basis functions for each data set to achieve an error less than 0.001\%. Figure \ref{fig:Kr_EIM} shows the results of the EIM approximation. We observe that the bases of $\epsilon_{LJ}$ and $\sigma_{LJ}$ coincide with the high likelihood values in the domain. This is consistent with our intuition that important regions of the likelihood should be included in the online estimation for better accuracy. 

\begin{figure}[h]
	\centering
	\begin{subfigure}[b]{0.32\textwidth}
		\centering
		\includegraphics[width=\textwidth]{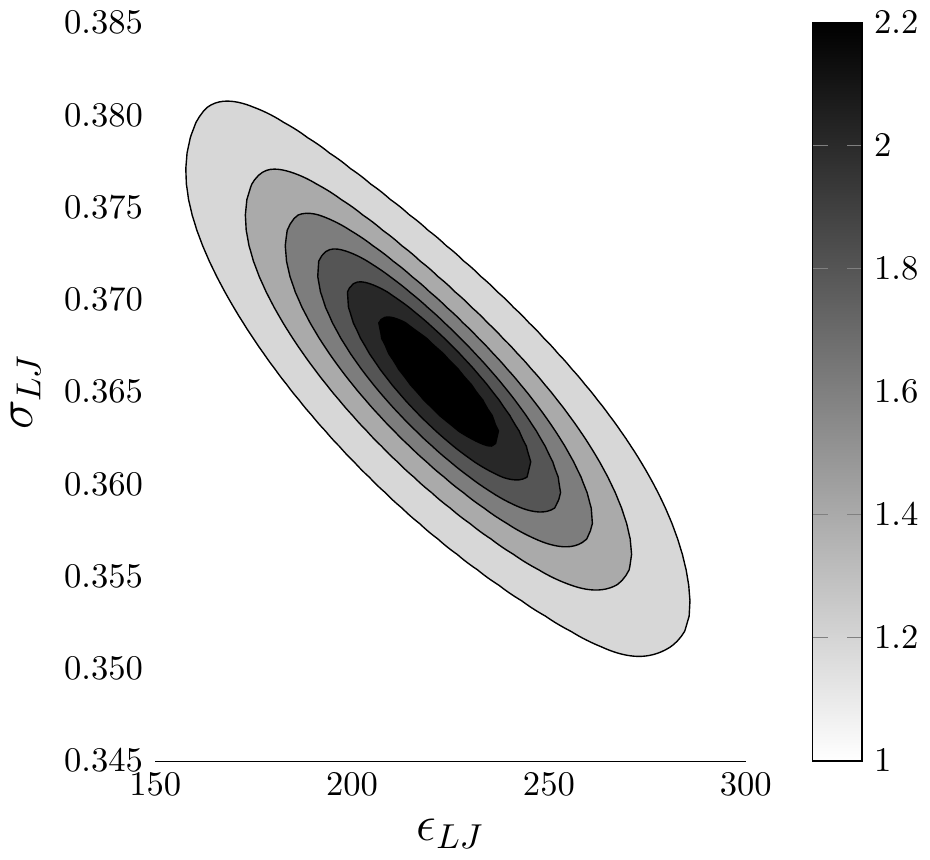}
		\caption{Posterior of $\epsilon_{LJ}$ and $\sigma_{LJ}$}
		\label{fig:Kr_para}
	\end{subfigure}
	\begin{subfigure}[b]{0.32\textwidth}
		\centering
		\includegraphics[width=0.8\textwidth]{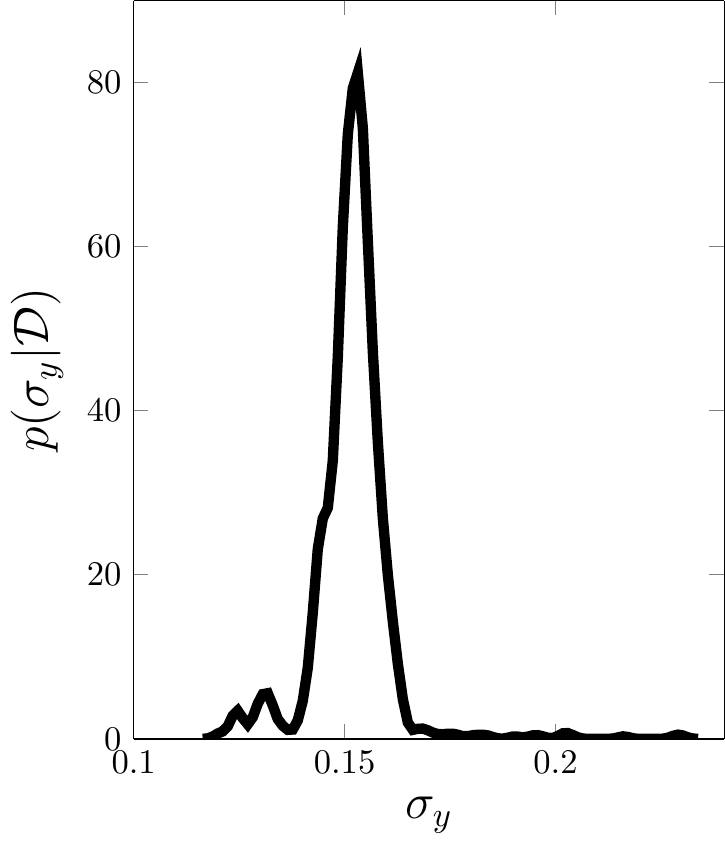}
		\caption{Posterior of $\sigma_y$}
		\label{fig:Kr_sigy}
	\end{subfigure}
	\begin{subfigure}[b]{0.32\textwidth}
		\centering
		\includegraphics[width=0.85\textwidth]{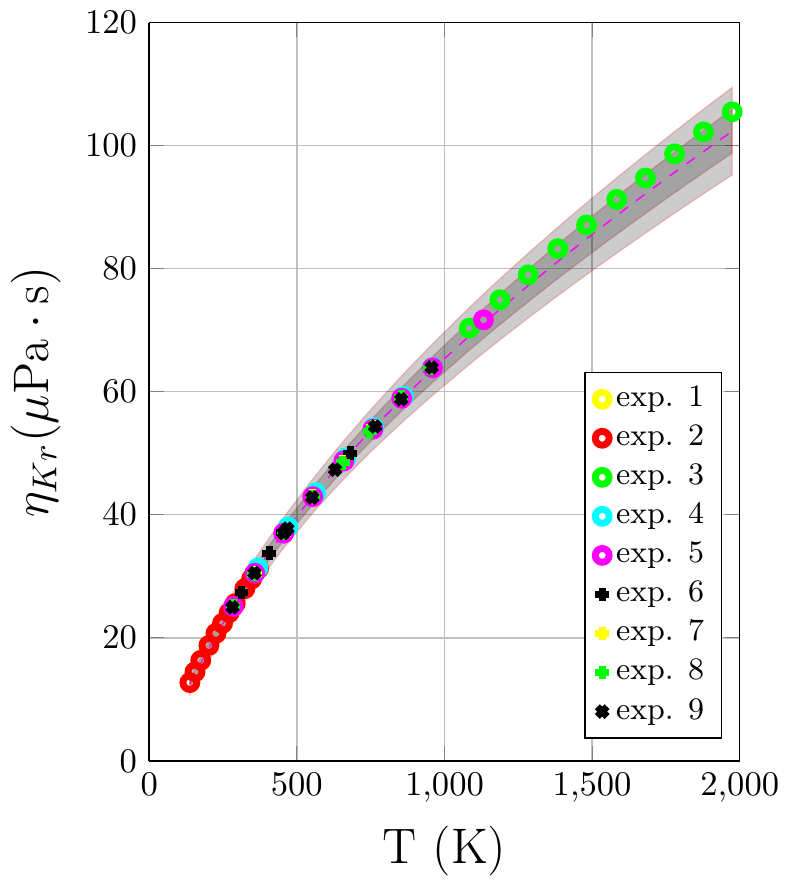}
		\caption{Posterior-robust prediction}
		\label{fig:Kr_pred}
	\end{subfigure}
		\caption{Posterior results for the Krypton MD simulation study.}
		\label{fig:Kr_fin}
\end{figure}

For each EIM basis, 2500 posterior samples of $\epsilon_{LJ}$ and $\sigma_{LJ}$ are recorded to be used in the post-processing step of our HSM analysis. We use BASIS to draw the samples because it is highly parallel and the evidence is a by-product of the algorithm \cite{Wu_BASIS:2016}. Then, we draw the posterior samples of the hyperparameters $\vpsi$ with the post-processing step in our method. BASIS is once again used to draw 1000 posterior samples of $\vpsi$. The posterior distribution of $\epsilon_{LJ}$ and $\sigma_{LJ}$ is approximated by $N$ posterior samples of $\vpsi$:
\begin{equation}
\begin{aligned}
	p(\epsilon_{LJ},\sigma_{LJ}|\mD) &= \int \! p(\epsilon_{LJ},\sigma_{LJ}|\vpsi)p(\vpsi|\mD)\,\mathrm{d} \vpsi\\
	&\approx \frac{1}{N}\sum_{i=1}^{N} p(\epsilon_{LJ},\sigma_{LJ}|\vpsi^{(i)}),~~\text{where}~~\vpsi^{(i)} \sim p(\vpsi|\mD)
\end{aligned}	
\end{equation}
Figure \ref{fig:Kr_fin} shows the results of the HSM analysis. We observe that our results are consistent with the results reported in \cite{Wu+et_al:2015}. This provides a verification of the accuracy of our method.

\subsection{PKPD model: Cancer drug}
Finley et al. \cite{Finley+et_al:2015} used the classical Bayesian model to study a PKPD model for the anti-vascular endothelial growth factor (anti-VEGF) cancer therapeutic agent, aflibercept. Multiple models with different number of parameters were calibrated using two types of clinical data: the plasma concentrations of free aflibercept and bound aflibercept. The data can be separated into 6 groups, each corresponding to a different dosage of the drug. We apply the HSM structure in Figure \ref{fig:BN_HB} to the basic model used in \cite{Finley+et_al:2015}, which has three PK model parameters and five prediction error parameters. The prediction of the two types of data using these parameters are calculated by first running the system to reach steady state, and then imposing the corresponding drug dosages. The three model parameters $\{k_{EC},k_{N},k_{T}\}$ represent the secretion rate of VEGF in the blood, normal tissue and tumor compartment, respectively. The five prediction error parameters $\{\sigma_{SS},\sigma_{BAs},\sigma_{BAf},\sigma_{FAs},\sigma_{FAf}\}$  represent standard deviations of zero-mean Gaussian distributions for five different types of prediction errors: $\sigma_{SS}$ --- error on the steady state prediction, $\sigma_{BAs}$ --- error on the bound aflibercept prediction scaled by time, $\sigma_{BAf}$ --- error on the bound aflibercept prediction without any scaling, $\sigma_{FAs}$ --- error on the free aflibercept prediction scaled by time, $\sigma_{FAf}$ --- error on the free aflibercept prediction without any scaling. We choose independent Gaussian prior distributions for all of the eight parameters in log-10 scale. 

\begin{figure}[h]
	\centering
	\begin{subfigure}[b]{0.45\textwidth}
		\centering
		\includegraphics[width=\textwidth]{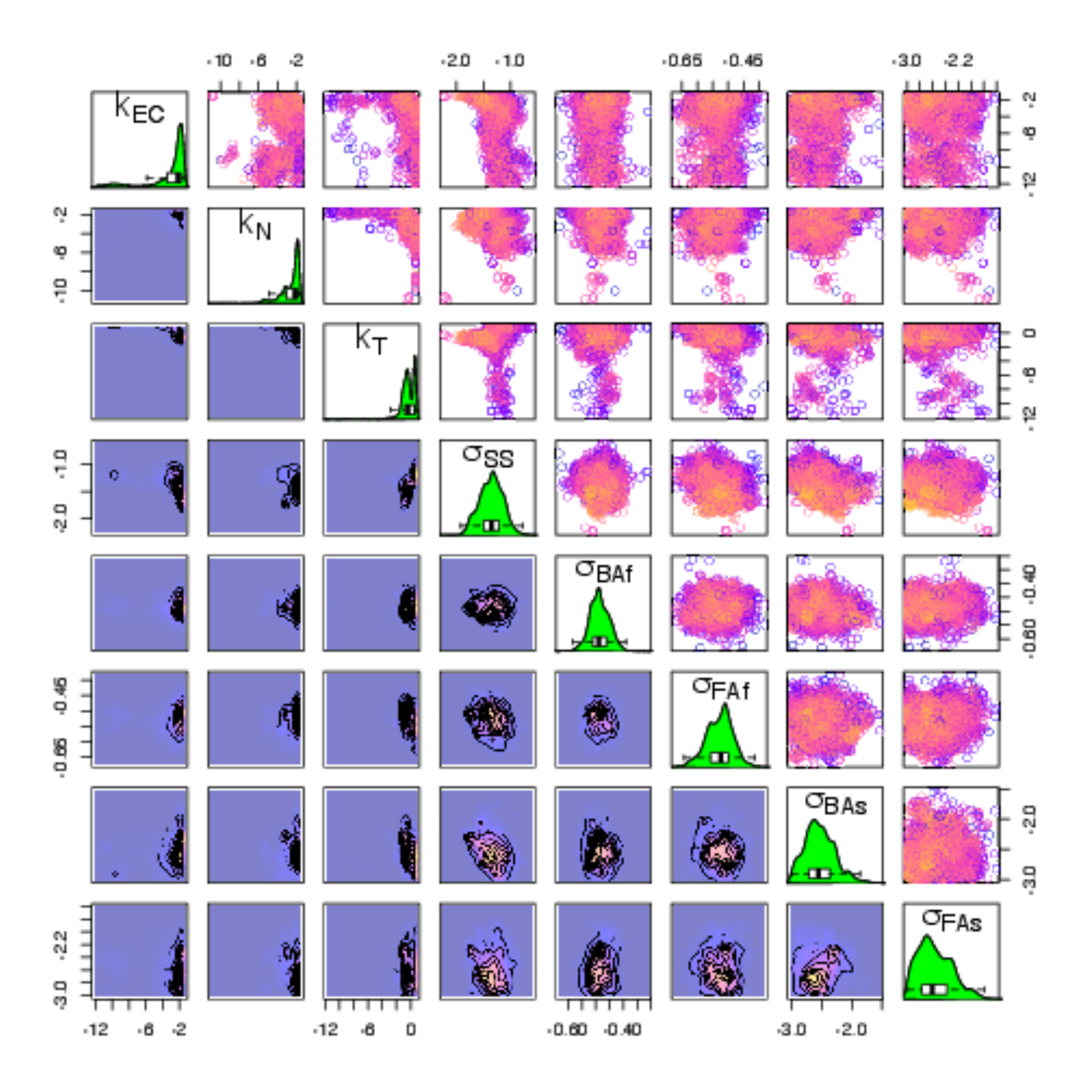}
		\caption{Classical Bayesian}
		\label{fig:PK_NB}
	\end{subfigure}
	\begin{subfigure}[b]{0.45\textwidth}
		\centering
		\includegraphics[width=\textwidth]{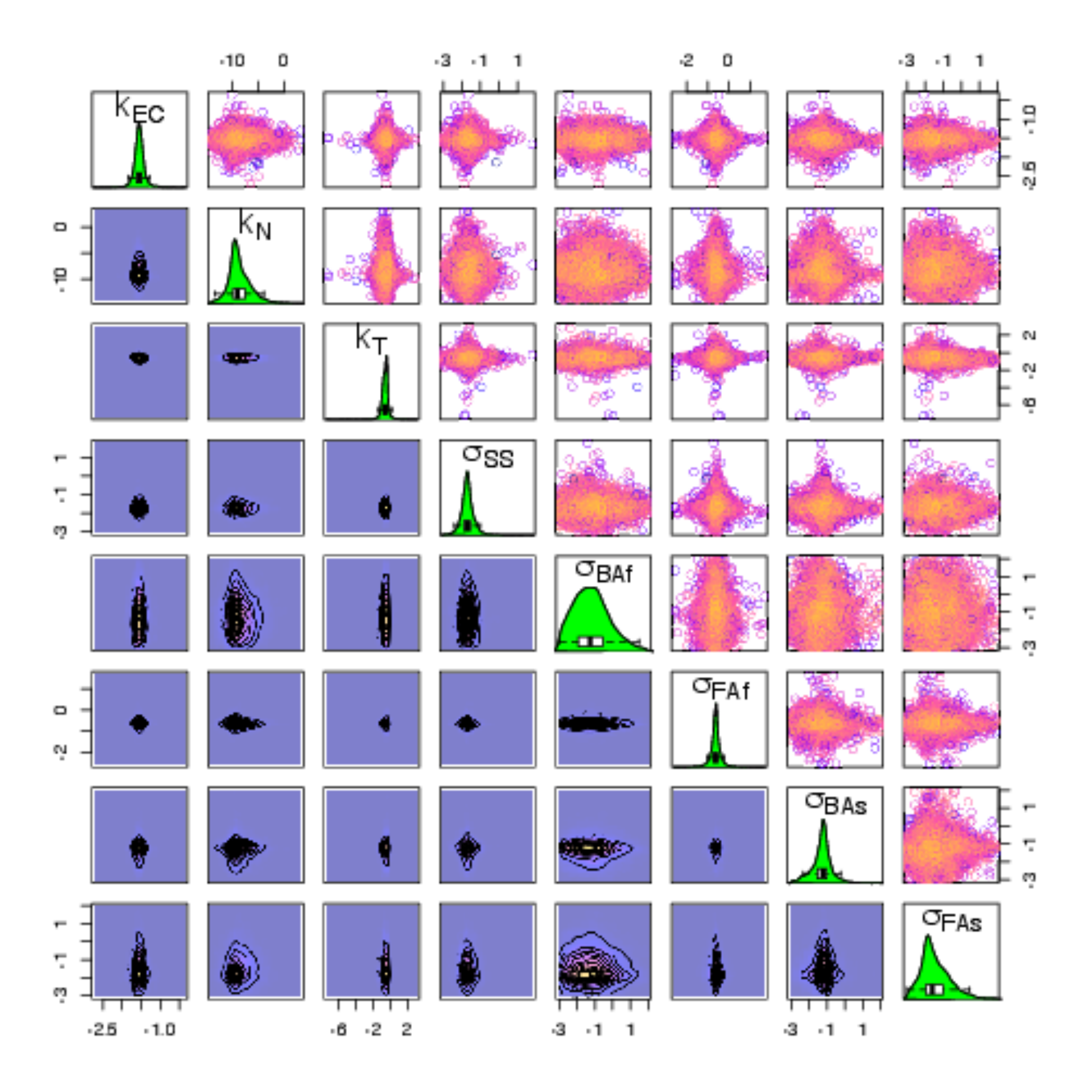}
		\caption{Basic HSM}
		\label{fig:PK_HB}
	\end{subfigure}
	\caption{Posterior distributions for parameters of the basic PK model. Upper diagonal: projection of the posterior samples for all pairs of 2D parameter space (colors indicate log-likelihood values of the samples). Diagonal: marginal distributions of the model parameters estimated using kernel histograms. Box-plots denote the means and the 5 and 95 percentiles. Lower diagonal: projected densities in 2D parameter space constructed via a kernel estimate (coloring according to log-posterior values).}
	\label{fig:PK_post}
\end{figure}

\begin{figure}[h]
	\centering
	\begin{subfigure}[b]{0.4\textwidth}
		\centering
		\includegraphics[trim=1cm 6cm 1cm 6cm,width=\textwidth]{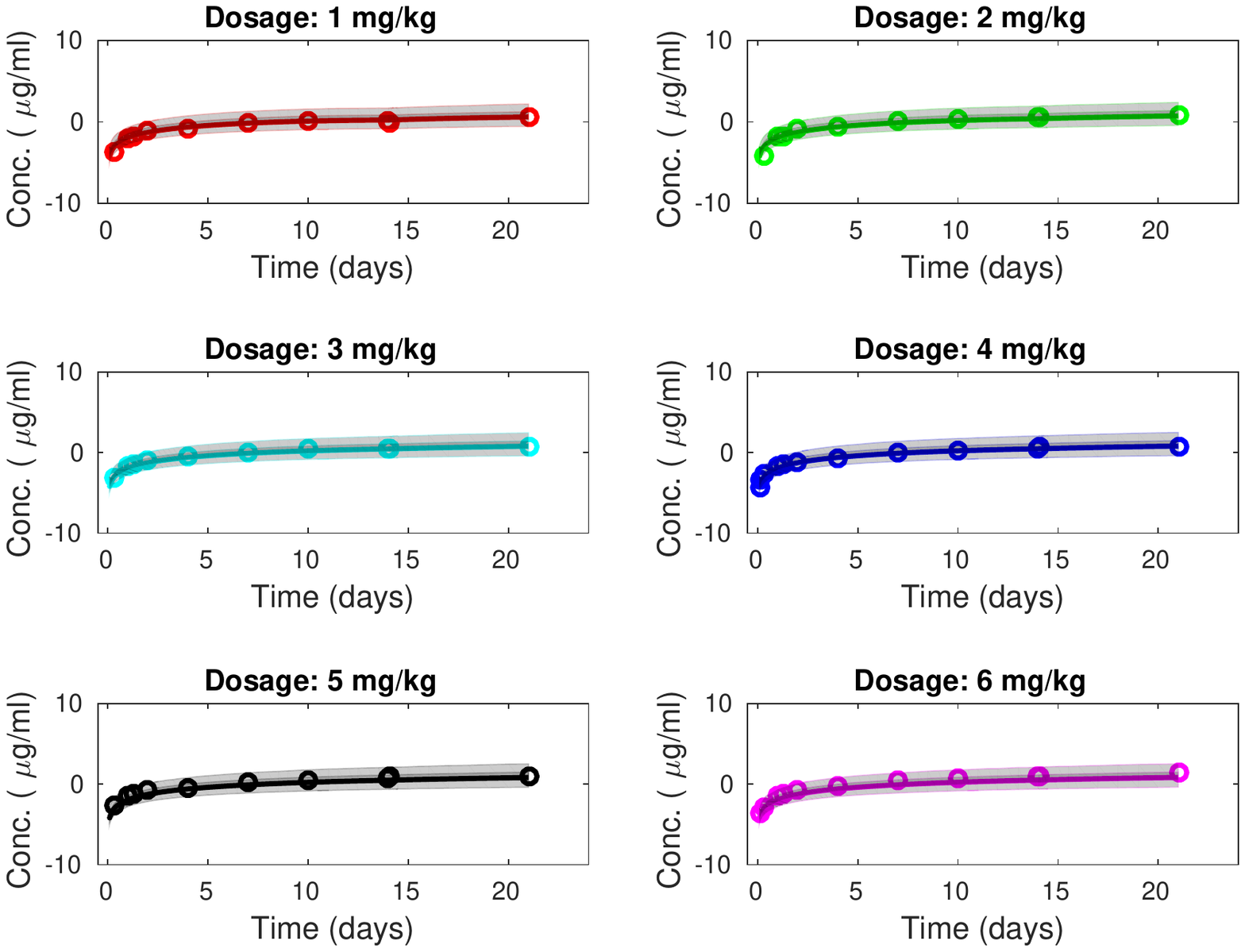}
		\caption{Bound aflibercept}
		\label{fig:VEGF_bd}
	\end{subfigure}
	\begin{subfigure}[b]{0.4\textwidth}
		\centering
		\includegraphics[trim=1cm 6cm 1cm 6cm,width=\textwidth]{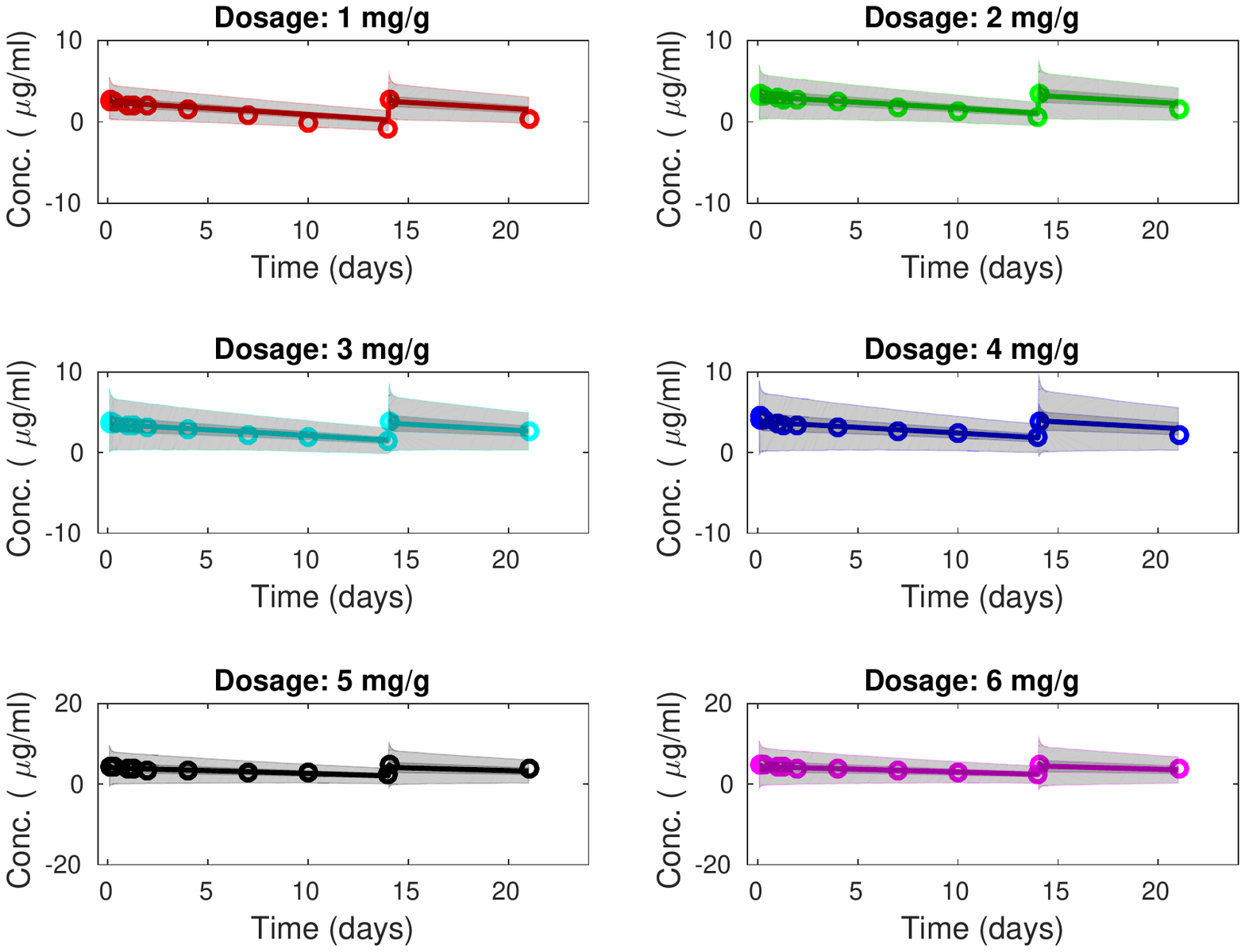}
		\caption{Free aflibercept}
		\label{fig:VEGF_free}
	\end{subfigure}
	\caption{Posterior robust predictions for the plasma concentrations of aflibercept based on the basic HSM. Circles are the data points. Dark gray and light gray region are the 50\% and 90\% quantile range of the posterior distribution, respectively.}
	\label{fig:VEGF_fin}
\end{figure}

Figure \ref{fig:PK_post} and Table \ref{tab:PK} compare the posterior distribution of the classical Bayesian model used in \cite{Finley+et_al:2015} with the posterior distribution of the basic HSM presented in this paper. We observe that fitting all data at once (the classical Bayesian model) leads to a significantly larger posterior variance for the model parameters, while the HSM leads to a larger posterior variance for the prediction error parameters. This implies that the knowledge about the model parameters is more transferable across different dosages than the knowledge about the prediction error parameters. We note that the steady state prediction is not affected by the change of dosage. Indeed, we observe that the inferred value of $\sigma_{SS}$ in the HSM is similar to the value for the classical Bayesian model. The slight increase of the CV of $\sigma_{SS}$ for the HSM case is expected because the data used for inference in the HSM is divided into groups of smaller data sets. Moreover, the scaled standard deviations $\sigma_{BAs}$ and $\sigma_{FAs}$ are larger, and the fixed standard deviations $\sigma_{BAf}$ and $\sigma_{FAf}$ are smaller in the HSM. This is also expected because in the classical Bayesian model, predictions for different dosages may have different level of sensitivity to the errors scaled by time. Hence, the classical Bayesian model tends to have a larger $\sigma_{BAf}$ and $\sigma_{FAf}$ than $\sigma_{BAs}$ and $\sigma_{FAs}$. As a result, the posterior robust prediction based on the HSM (Figure \ref{fig:VEGF_fin}) has completely different prediction errors for both bound aflibercept and free aflibercept compared to the one in \cite{Finley+et_al:2015}.

\begin{table}
	\footnotesize
	\caption{Posterior sample mean and coefficient of variation (CV) for parameters of the PK model. All samples are drawn in log-space. The mean is converted to the linear scale. The CV is calculated based on the samples in log-scale.}
	\label{tab:PK}
	\centering
	\begin{tabular}{|c|c|c|c|c|}
		\hline
		& \multicolumn{2}{c|}{Classical Bayesian} & \multicolumn{2}{c|}{Basic HSM} \\ \cline{2-5}
		& Mean & CV(\%) & Mean & CV(\%) \\ \hline
		$k_{EC}$ & 0.035 & 191.7 & 0.212 & 1.2 \\ \hline
		$k_{N}$ & 0.073 & 43.5 & $2\times10^{-4}$ & 52.0 \\ \hline
		$k_{T}$ & 0.634 & 442.7 & 0.535 & 23.6 \\ \hline
		$\sigma_{SS}$ & 0.257 & 3.4 & 0.183 & 7.5 \\ \hline
		$\sigma_{BAf}$ & 0.615 & 0.3 & 0.302 & 81.9 \\ \hline
		$\sigma_{FAf}$ & 0.590 & 0.3 & 0.534 & 11.0 \\ \hline
		$\sigma_{BAs}$ & 0.080 & 2.4 & 0.276 & 26.2 \\ \hline
		$\sigma_{FAs}$ & 0.077 & 2.7 & 0.235 & 44.2 \\ \hline
	\end{tabular}
\end{table}

\section{Conclusion}\label{sec:con}
We demonstrate the benefits of using the hierarchical Bayesian framework in complex systems. Because of the double usage of the term HBM in the literature, we first explain the distinction between two very different HBMs: the HPM and the HSM. Then, we focus on studying the HSM, which has many interesting theoretical implications, as well as many computational challenges in practice. Based on examples of polynomial functions, we suggest that the HSM is capable of explicitly separating different types of uncertainties in a system, and can be an effective tool for modeling uncertainty of reduced order models. In order to apply HSM to practical problems, we propose an efficient approximation method to tackle the high computationally cost associated with using HSMs. Our method is a ``bottom-up" approach that begins by drawing posterior samples of the model parameters for each data set. Then, we use a post-processing step to move up the hierarchy of the parameters for drawing posterior samples. Building on the basic HSM, we demonstrate our method using two alternative HSMs that have a more complicated hierarchical structure. Lastly, we validate our method using two illustrative examples based on previous studies: a molecular dynamics simulation of Krypton and a PKPD model of a cancer drug.

The HSM is a convenient and effective tool to build the stochastic model/likelihood for a complicated system. It also opens up new types of analyses and it results in different conclusions as compared to classical Bayesian inference. Our approximation method provides a standard procedure for analyzing hierarchical models. Beginning with an analysis for each individual data set, our method allows us to move up the hierarchy of the parameters efficiently to potentially extract more in-depth information about the system of interest. Moreover, it is very efficient for sequentially received data sets because our estimation scheme is a computationally fast post-processing step. So far, we have mainly used a purely statistical (empirical) model for $p(\vtheta|\vpsi)$. In future work, we plan to employ other choices for modeling $p(\vtheta|\vpsi)$, which may motivate the development of new modeling tools.

\begin{appendices}
\section{Derivations}\label{app:A}
We consider a set of data $\mD$ structured as $\mD = \{D_i | i = 1,\dots, N_D\}$ with each subset of data $D_i = \{(x_{j,i},y_{j,i}) | j = 1,\dots,N_{D_i}\}$, where $(x_{j,i},y_{j,i})$ denotes a single data point. Using vector notations, we denote $\vec{x}_i = (x_{1,i}, \dots , x_{N_{D_i},i})^T$, $\vec{y}_i = (y_{1,i}, \dots , y_{N_{D_i},i})^T$, $\vec{x} = (\vec{x}_1^T, \dots , \vec{x}_{N_D}^T)^T$ and $\vec{y} = (\vec{y}_1^T, \dots , \vec{y}_{N_D}^T)^T$. Also, we denote the total number of data points $N_d = \sum_{i=1}^{N_D} N_{D_i}$. Hence, $\vec{x}$ and $\vec{y}$ are vectors with $N_d$ elements.

Starting with a general formulation, we denote $\vec{\theta}$ as a vector of all model parameters and $\vec{\psi}$ as a vector of all hyperparameters. Repeatedly using Bayes' Theorem and the total probability theorem, we can derive the following expressions for a given model $\mM_k$:
\begin{align}
	p(\vec{\psi}|\mD,\mM_k) &= \frac{p(\mD|\vec{\psi},\mM_k)p(\vec{\psi}|\mM_k)}{p(\mD|\mM_k)} \label{eq:A1}\\
	p(\vec{\theta}|\mD,\mM_k) &= \int \! p(\vec{\theta}|\mD,\vec{\psi},\mM_k)p(\vec{\psi}|\mD,\mM_k)\, \mathrm{d}\vec{\psi} \label{eq:A2}\\
	p(\vec{\theta}|\mD,\vec{\psi},\mM_k) &= \frac{p(\mD|\vec{\theta},\vec{\psi},\mM_k)p(\vec{\theta}|\vec{\psi},\mM_k)}{p(\mD|\vec{\psi},\mM_k)} \label{eq:A3}\\
	p(\mM_k|\mD) &= \frac{p(\mD|\mM_k)p(\mM_k)}{p(\mD)} \label{eq:A4}\\
	p(\mD|\mM_k) &= \int \! p(\mD|\vec{\psi},\mM_k)p(\vec{\psi}|\mM_k)\, \mathrm{d}\vec{\psi} \label{eq:A5}\\
	p(\mD|\vec{\psi},\mM_k) &= \int \! p(\mD|\vec{\theta},\vec{\psi},\mM_k)p(\vec{\theta}|\vec{\psi},\mM_k)\, \mathrm{d}\vec{\theta} \label{eq:A6}
\end{align}
We note that usually $p(\mM_k)$ is constant for all $k$ because we do not want to introduce bias to any model before data is available. Hence, $p(\mM_k|\mD) \propto p(\mD|\mM_k)$. Also, $\vec{\theta}$ is simply a scalar $\theta$ in our study and the likelihood standard deviation $\sigma_y$ is inferred separately. In the following, we derive the analytical expressions for the posterior distributions, the evidences and the robust-posterior predictions of the different models.

% Model 1a
\subsection{Non-hierarchical model, $\mM_{1a}$}
In this model, all data points are independent when $\theta$ and $\sigma_y$ are known. By assuming a Gaussian distribution for the likelihood and a uniform prior $U(\theta) = 1/c_\theta$, we can derive that:
\begin{equation}\label{eq:likeT}
	\begin{aligned}
	p(\mathcal{D}|\theta,\sigma_y,\mM_{1a}) &= \prod_{i=1}^{N_D}\prod_{j=1}^{N_{D_i}} \frac{1}{\sqrt{2\pi} \sigma_y} \mathrm{exp}\left(-\frac{1}{2\sigma_y^2} (y_{j,i} - \theta x_{j,i})^2 \right) \\
%&= (2\pi)^{-N_d/2} \sigma_y^{-N_d} \mathrm{exp} \left( -\frac{1}{2\sigma_y^2} \sum_{i=1}^{N_d} (y_i - \theta x_i)^2 \right) \\
	&= (2\pi)^{-N_d/2} \sigma_y^{-N_d} \mathrm{exp} \left( -\frac{1}{2\sigma_y^2} (\vec{y} - \theta \vec{x})^T (\vec{y} - \theta \vec{x}) \right)
	\end{aligned}
\end{equation}
%\begin{equation}\label{eq:lnlikeT}
%	\begin{aligned}
%	\mathrm{ln} p(\mathcal{D}|\theta,\sigma_y) &= -\frac{N_d}{2} \mathrm{ln}(2\pi) -N_d \mathrm{ln}\sigma_y -\frac{1}{2\sigma_y^2} \sum_{i=1}^{N_d} (y_i - \theta x_i)^2 \\
%	&= -\frac{N_d}{2} \mathrm{ln}(2\pi) -N_d \mathrm{ln}\sigma_y -\frac{1}{2\sigma_y^2} \left(\vec{y}^T \vec{y} - 2\theta \vec{x}^T \vec{y} + \theta^2 \vec{x}^T \vec{x} \right)
%	\end{aligned}
%\end{equation}
Using Bayes' theorem, the posterior distribution is proportional to only the likelihood when the prior is uniform. It can be obtained from completing square for the exponential part of the Gaussian likelihood:
\begin{equation}
	\begin{aligned}
%	-\frac{1}{2\sigma_y^2} \left(\vec{y}^T \vec{y} - 2\theta \vec{x}^T \vec{y} + \theta^2 \vec{x}^T \vec{x} \right) &= -\frac{\vec{x}^T \vec{x}}{2\sigma_y^2} \left( \frac{\vec{y}^T \vec{y}}{\vec{x}^T \vec{x}} - 2\theta \frac{\vec{x}^T \vec{y}}{\vec{x}^T \vec{x}} + \theta^2 \right) \\
	-\frac{1}{2\sigma_y^2} (\vec{y} - \theta \vec{x})^T (\vec{y} - \theta \vec{x}) &= -\frac{\vec{x}^T \vec{x}}{2\sigma_y^2} \left( \frac{\vec{y}^T \vec{y}}{\vec{x}^T \vec{x}} - 2\theta \frac{\vec{x}^T \vec{y}}{\vec{x}^T \vec{x}} + \theta^2 \right) \\
	&= -\frac{1}{2\tilde{\sigma}_{\theta}^2} \left( (\theta - \tilde{\mu}_{\theta})^2 + \frac{\vec{y}^T \vec{y}}{\vec{x}^T \vec{x}} - \tilde{\mu}_{\theta}^2\right) \\
	\text{where  } &\tilde{\sigma}_{\theta}^2 = \frac{\sigma_y^2}{\vec{x}^T \vec{x}}, \, \tilde{\mu}_{\theta} = \frac{\vec{x}^T \vec{y}}{\vec{x}^T \vec{x}} = \frac{\vec{x}^T \vec{y}}{\sigma_y^2}\tilde{\sigma}_{\theta}^2
	\end{aligned}
\end{equation}
\begin{equation}
	p(\mD|\theta,\sigma_y,\mM_{1a}) = \frac{1}{\sqrt{2\pi} \tilde{\sigma}_{\theta}}\mathrm{exp}\left(-\frac{(\theta - \tilde{\mu}_{\theta})^2}{2 \tilde{\sigma}_{\theta}^2} \right) \frac{\tilde{\sigma}_{\theta}}{(2\pi)^{\frac{N_d-1}{2}} \sigma_y^{N_d}}\mathrm{exp}\left(-\frac{\frac{\vec{y}^T \vec{y}}{\vec{x}^T \vec{x}} - \tilde{\mu}_{\theta}^2}{2 \tilde{\sigma}_{\theta}^2} \right)
\end{equation}
Hence, the posterior distribution is also Gaussian:
\begin{equation}\label{eq:thetaP}
	p(\theta|\mD,\sigma_y,\mM_{1a}) = N(\theta|\tilde{\mu}_{\theta},\tilde{\sigma}_{\theta}^2), \,\, \tilde{\sigma}_{\theta} = \frac{\sigma_y}{\sqrt{\vec{x}^T \vec{x}}}, \, \tilde{\mu}_{\theta} = \frac{\vec{x}^T \vec{y}}{\vec{x}^T \vec{x}} = \frac{\vec{x}^T \vec{y}}{\sigma_y^2}\tilde{\sigma}_{\theta}^2
\end{equation}
and the evidence term $p(\mD|\sigma_y,\mM_{1a})$ calculated using Equation \ref{eq:A6} (substitute $\vpsi$ by $\sigma_y$) is:
\begin{equation}
	p(\mD|\sigma_y,\mM_{1a}) = \frac{1}{c_{\theta}}\frac{\tilde{\sigma}_{\theta}}{(2\pi)^{\frac{N_d-1}{2}} \sigma_y^{N_d}}\mathrm{exp}\left(-\frac{1}{2} \left( \frac{\vec{y}^T \vec{y}}{\sigma_y^2} - \frac{\tilde{\mu}_{\theta}^2}{\tilde{\sigma}_{\theta}^2} \right) \right)
\end{equation}
Then, we use $N_s$ samples from Monte Carlo Simulation to estimate the posterior of $\sigma_y$ and the model evidence:
\begin{equation}\label{eq:sigy_M1}
	\begin{aligned}
	&p(\sigma_y|\mD,\mM_{1a}) \approx \sum_{j=1}^{N_s} w_j \delta(\sigma_y - \sigma_y^{(j)})\\
	\text{where} \,\,\,\,\, &\sigma_y^{(j)} \sim p(\sigma_y|\mM_{1a}), w_j \propto p(\mD|\sigma_y^{(j)},\mM_{1a}), \sum_{j=1}^{N_s} w_j = 1
	\end{aligned}
\end{equation}
\begin{equation}
	p(\mD|\mM_{1a}) \approx \frac{1}{N_s} \sum_{j=1}^{N_s} p(\mD|\sigma_y^{(j)},\mM_{1a})
\end{equation}
The marginalized posterior of $\theta$ is estimated by substituting Equation \ref{eq:thetaP} and \ref{eq:sigy_M1} into Equation \ref{eq:A2}, which we substitute $\vpsi$ by $\sigma_y$:
\begin{equation}
	\begin{aligned}
	p(\theta|\mD,\mM_{1a}) &\approx \sum_{j=1}^{N_s} w_j p(\theta|\mD,\sigma_y^{(j)},\mM_{1a})\\
	&= \sum_{j=1}^{N_s} w_j N(\theta|\tilde{\mu}_{\theta},(\tilde{\sigma}_{\theta}^{(j)})^2)\\
	\text{where} \,\,\,\,\, &\tilde{\mu}_{\theta} = \frac{\vec{x}^T \vec{y}}{\vec{x}^T \vec{x}}, \,\,\, \tilde{\sigma}_{\theta}^{(j)} = \frac{\sigma_y^{(j)}}{\sqrt{\vec{x}^T \vec{x}}} 
	\end{aligned}
\end{equation}
As a result, the statistics of the marginalized posterior of $\theta$ and $\sigma_y$ can also be estimated ($E[\cdot|\mD]$ --- posterior mean; $Std[\cdot|\mD]$ --- posterior standard deviation) based on the MCS samples:
\begin{equation}
	\begin{aligned}
	E[\theta|\mD] &= \int \! \theta p(\theta|\mD,\mM_{1a}) \, \mathrm{d} \theta \approx \tilde{\mu}_\theta\\
	Std[\theta|\mD] &\approx \sqrt{\left( \sum_{j=1}^{N_s} w_j \left( (\tilde{\sigma_\theta}^{(j)})^2 + \tilde{\mu}_\theta^2 \right)^2 \right) - E[\theta|\mD]^2}\\
	E[\sigma_y|\mD] &\approx \sum_{j=1}^{N_s} w_j \sigma_y^{(j)}\\
	Std[\sigma_y|\mD] &= \sqrt{\left( \sum_{j=1}^{N_s} w_j (\sigma_y^{(j)})^2 \right) - E[\sigma_y|\mD]^2}
	\end{aligned}
\end{equation}
To perform robust prediction of a new point $(\hat{x},\hat{y})$, we need to evaluate $p((\hat{x},\hat{y})|\mD,\mM_{1a})$. Again, we can use the posterior samples from Equation \ref{eq:sigy_M1}:
\begin{equation}
	\begin{aligned}
	p((\hat{x},\hat{y})|\mD,\mM_{1a}) &= \int \! p((\hat{x},\hat{y})|\sigma_y,\mD,\mM_{1a})p(\sigma_y|\mD,\mM_{1a}) \, \mathrm{d}\sigma_y\\
	&\approx \sum_{j=1}^{N_s} w_j p((\hat{x},\hat{y})|\sigma_y^{(j)},\mD,\mM_{1a})
	\end{aligned}
\end{equation}
where $p((\hat{x},\hat{y})|\sigma_y^{(j)},\mD,\mM_{1a})$ can be evaluated analytically based on the properties of the product of two Gaussian distributions (applied to the last line of this equation):
\begin{equation}
	\begin{aligned}
	&p((\hat{x},\hat{y})|\sigma_y^{(j)},\mD,\mM_{1a})\\ 
	=& \int \! p((\hat{x},\hat{y})|\theta,\sigma_y^{(j)},\mM_{1a})p(\theta|\sigma_y^{(j)},\mD,\mM_{1a}) \, \mathrm{d}\theta\\
	=& \int \! \frac{1}{\sqrt{2\pi}\sigma_y^{(j)}}\mathrm{exp}\left( -\frac{(\hat{y} - \theta \hat{x})^2}{2(\sigma_y^{(j)})^2} \right) \frac{1}{\sqrt{2\pi}\tilde{\sigma}_\theta^{(j)}}\mathrm{exp}\left( -\frac{(\theta - \tilde{\mu}_\theta)^2}{2(\tilde{\sigma}_\theta^{(j)})^2} \right) \, \mathrm{d}\theta\\
	=& \int \! \frac{1}{\hat{x}\sqrt{2\pi}(\sigma_y^{(j)}/\hat{x})}\mathrm{exp}\left( -\frac{(\theta - \hat{y}/\hat{x})^2}{2(\sigma_y^{(j)} / \hat{x})^2} \right) \frac{1}{\sqrt{2\pi}\tilde{\sigma}_\theta^{(j)}}\mathrm{exp}\left( -\frac{(\theta - \tilde{\mu}_\theta)^2}{2(\tilde{\sigma}_\theta^{(j)})^2} \right) \, \mathrm{d}\theta\\
	=& \frac{1}{\hat{x}\sqrt{2\pi \left( (\sigma_y^{(j)}/\hat{x})^2 + (\tilde{\sigma}_\theta^{(j)})^2 \right)}}\mathrm{exp}\left( -\frac{(\tilde{\mu}_\theta - \hat{y}/\hat{x})^2}{2\left((\sigma_y^{(j)}/\hat{x})^2 + (\tilde{\sigma}_\theta^{(j)})^2\right)} \right)
	\end{aligned}
\end{equation}

% Model 1b
\subsection{HPM, $\mM_{1b}$}\label{sec:A2}
As explained in Section \ref{sec:HPM}, this model is essentially the same as $\mM_{1a}$, but with a different prior of $\theta$. By assuming a Gaussian distribution for the likelihood and the prior of $\theta$ given the hyperparameters $\vpsi = \{\mu_\theta,\sigma_\theta\}$, we can derive that:
\begin{equation}
\begin{aligned}
&p(\mD|\theta,\sigma_y,\mM_{1b})p(\theta|\mu_\theta,\sigma_\theta,\mM_{1b})\\ 
= &\frac{1}{(2\pi)^{N_d/2} \sigma_y^{N_d}} \mathrm{exp} \left( -\frac{(\vec{y} - \theta \vec{x})^T (\vec{y} - \theta \vec{x})}{2\sigma_y^2} \right) \frac{1}{\sqrt{2\pi}\sigma_{\theta}} \mathrm{exp} \left( -\frac{(\theta - \mu_{\theta})^2}{2\sigma_{\theta}^2} \right)
\end{aligned}
\end{equation}
Using Bayes' theorem, the posterior distribution is proportional to the product of the likelihood and the prior. It can be obtained from completing square for the exponential part of the Gaussian distributions:
\begin{equation}
\begin{aligned}
&-\frac{1}{2\sigma_y^2} \left(\vec{y}^T \vec{y} - 2\theta \vec{x}^T \vec{y} + \theta^2 \vec{x}^T \vec{x} \right) -\frac{1}{2\sigma_{\theta}^2} \left( \theta^2 - 2\theta \mu_{\theta} + \mu_{\theta}^2 \right) \\
= &-\frac{1}{2} \left( \theta^2 \left(\frac{\vec{x}^T \vec{x}}{\sigma_y^2}+\frac{1}{\sigma_{\theta}^2} \right) - 2\theta \left( \frac{\vec{x}^T \vec{y}}{\sigma_y^2}+\frac{\mu_{\theta}}{\sigma_{\theta}^2} \right) + \frac{\vec{y}^T \vec{y}}{\sigma_y^2} + \frac{\mu_{\theta}^2}{\sigma_{\theta}^2}\right) \\
= &-\frac{1}{2\tilde{\sigma}_{\theta}^2} \left( \theta^2 - 2\theta \left( \frac{\vec{x}^T \vec{y} \sigma_{\theta}^2 + \mu_{\theta}\sigma_y^2}{\vec{x}^T \vec{x}\sigma_{\theta}^2+\sigma_y^2} \right) + \frac{\vec{y}^T \vec{y}\sigma_{\theta}^2 + \mu_{\theta}^2\sigma_y^2}{\vec{x}^T \vec{x}\sigma_{\theta}^2+\sigma_y^2} \right) \\
= &-\frac{1}{2\tilde{\sigma}_{\theta}^2} \left( \left(\theta - \tilde{\mu}_{\theta}\right)^2 + \frac{\vec{y}^T \vec{y}\sigma_{\theta}^2 + \mu_{\theta}^2\sigma_y^2}{\vec{x}^T \vec{x}\sigma_{\theta}^2+\sigma_y^2} - \tilde{\mu}_{\theta}^2 \right) \\
&\text{where  } \tilde{\sigma}_{\theta}^2 = \frac{\sigma_{\theta}^2\sigma_y^2}{\vec{x}^T \vec{x}\sigma_{\theta}^2+\sigma_y^2}, \, \tilde{\mu}_{\theta} = \frac{\vec{x}^T \vec{y} \sigma_{\theta}^2 + \mu_{\theta}\sigma_y^2}{\vec{x}^T \vec{x}\sigma_{\theta}^2+\sigma_y^2} = \left( \frac{\vec{x}^T \vec{y}}{\sigma_y^2}+\frac{\mu_{\theta}}{\sigma_{\theta}^2} \right)\tilde{\sigma}_{\theta}^2
\end{aligned}
\end{equation}
\begin{equation}
\begin{aligned}
&p(\mD|\theta,\sigma_y,\mM_{1b})p(\theta|\mu_\theta,\sigma_\theta,\mM_{1b}) \\
= &\frac{1}{\sqrt{2\pi} \tilde{\sigma}_{\theta}}\mathrm{exp}\left(-\frac{(\theta - \tilde{\mu}_{\theta})^2}{2 \tilde{\sigma}_{\theta}^2} \right) \frac{\tilde{\sigma}_{\theta}}{(2\pi)^{\frac{N_d}{2}} \sigma_y^{N_d}\sigma_{\theta}}\mathrm{exp}\left(-\frac{1}{2}\left( \frac{\vec{y}^T \vec{y}}{\sigma_y^2} + \frac{\mu_{\theta}^2}{\sigma_{\theta}^2} - \frac{\tilde{\mu}_{\theta}^2}{\tilde{\sigma}_{\theta}^2}\right) \right)
\end{aligned}
\end{equation}
Hence, the posterior distribution is also Gaussian:
\begin{equation}\label{eq:thetaPM2c}
\begin{aligned}
&p(\theta|\mD,\mu_\theta,\sigma_\theta,\sigma_y,\mM_{1b}) = N(\theta|\tilde{\mu}_{\theta},\tilde{\sigma}_{\theta}^2)\\
\text{where} \,\,\, &\tilde{\sigma}_{\theta} = \frac{\sigma_{\theta}\sigma_y}{\sqrt{\vec{x}^T \vec{x}\sigma_{\theta}^2+\sigma_y^2}}, \, \tilde{\mu}_{\theta} = \frac{\vec{x}^T \vec{y} \sigma_{\theta}^2 + \mu_{\theta}\sigma_y^2}{\vec{x}^T \vec{x}\sigma_{\theta}^2+\sigma_y^2} = \left( \frac{\vec{x}^T \vec{y}}{\sigma_y^2}+\frac{\mu_{\theta}}{\sigma_{\theta}^2} \right)\tilde{\sigma}_{\theta}^2
\end{aligned}
\end{equation}
and the evidence term $p(\mD|\mu_\theta,\sigma_\theta,\sigma_y,\mM_{1b})$ calculated using Equation \ref{eq:A6} is:
\begin{equation}
p(\mD|\mu_\theta,\sigma_\theta,\sigma_y,\mM_{1b}) = \frac{\tilde{\sigma}_{\theta}}{(2\pi)^{\frac{N_d}{2}} \sigma_y^{N_d}\sigma_{\theta}}\mathrm{exp}\left(-\frac{1}{2}\left( \frac{\vec{y}^T \vec{y}}{\sigma_y^2} + \frac{\mu_{\theta}^2}{\sigma_{\theta}^2} - \frac{\tilde{\mu}_{\theta}^2}{\tilde{\sigma}_{\theta}^2}\right) \right)
\end{equation}
We use $N_s$ samples from Monte Carlo Simulation to estimate the posterior of $\mu_\theta$, $\sigma_\theta$ and $\sigma_y$:
\begin{equation}\label{eq:sigy_M2c}
\begin{aligned}
p(\mu_\theta,\sigma_\theta,\sigma_y|\mD,\mM_{1b}) &\propto p(\mD|\mu_\theta,\sigma_\theta,\sigma_y,\mM_{1b})p(\mu_\theta,\sigma_\theta,\sigma_y|\mM_{1b})\\
&\approx \sum_{j=1}^{N_s} w_j \delta(\mu_\theta - \mu_\theta^{(j)}) \delta(\sigma_\theta - \sigma_\theta^{(j)})\delta(\sigma_y - \sigma_y^{(j)})
\end{aligned}
\end{equation}
\begin{equation*}
\text{where} \,\,\,\,\, (\mu_\theta^{(j)},\sigma_\theta^{(j)},\sigma_y^{(j)}) \sim p(\mu_\theta,\sigma_\theta,\sigma_y|\mM_{1b}), w_j \propto p(\mD|\mu_\theta^{(j)},\sigma_\theta^{(j)},\sigma_y^{(j)},\mM_{1b}), \sum_{j=1}^{N_s} w_j = 1
\end{equation*}
and the model evidence:
\begin{equation}
p(\mD|\mM_{1b}) \approx \frac{1}{N_s} \sum_{j=1}^{N_s} p(\mD|\mu_\theta^{(j)},\sigma_\theta^{(j)},\sigma_y^{(j)},\mM_{1b})
\end{equation}
The marginalized posterior of $\theta$ can be estimated by substituting Equation \ref{eq:thetaPM2c} and \ref{eq:sigy_M2c} into Equation \ref{eq:A2}:
\begin{equation}
\begin{aligned}
&p(\theta|\mD,\mM_{1b}) \approx \sum_{j=1}^{N_s} w_j N(\theta|\tilde{\mu}_{\theta}^{(j)},(\tilde{\sigma}_{\theta}^{(j)})^2)\\
\text{where} \,\,\, &\tilde{\sigma}_{\theta}^{(j)} = \frac{\sigma_{\theta}^{(j)}\sigma_y^{(j)}}{\sqrt{\vec{x}^T \vec{x}(\sigma_{\theta}^{(j)})^2+(\sigma_y^{(j)})^2}}, \, \tilde{\mu}_{\theta}^{(j)} = \frac{\vec{x}^T \vec{y} (\sigma_{\theta}^{(j)})^2 + \mu_{\theta}^{(j)}(\sigma_y^{(j)})^2}{\vec{x}^T \vec{x}(\sigma_{\theta}^{(j)})^2+(\sigma_y^{(j)})^2}
\end{aligned}
\end{equation}
As a result, the statistics of the marginalized posterior of $\theta$ and $\sigma_y$ can also be estimated based on the MCS samples:
\begin{equation}
\begin{aligned}
E[\theta|\mD] &= \int \! \theta p(\theta|\mD,\mM_{1b}) \, \mathrm{d} \theta \approx \sum_{j=1}^{N_s} w_j \tilde{\mu}_\theta^{(j)}\\
Std[\theta|\mD] &\approx \sqrt{\left( \sum_{j=1}^{N_s} w_j \left( (\tilde{\sigma}_\theta^{(j)})^2 + (\tilde{\mu}_\theta^{(j)})^2 \right)^2 \right) - E[\theta|\mD]^2}\\
E[\sigma_y|\mD] &\approx \sum_{j=1}^{N_s} w_j \sigma_y^{(j)}\\
Std[\sigma_y|\mD] &= \sqrt{\left( \sum_{j=1}^{N_s} w_j (\sigma_y^{(j)})^2 \right) - E[\sigma_y|\mD]^2}
\end{aligned}
\end{equation}
To perform robust prediction of a new point $(\hat{x},\hat{y})$, we need to evaluate $p((\hat{x},\hat{y})|\mD,\mM_{1b})$. Again, we can use the posterior samples from Equation \ref{eq:sigy_M2c}:
\begin{equation}
\begin{aligned}
p((\hat{x},\hat{y})|\mD,\mM_{1b}) &= \int \! p((\hat{x},\hat{y})|\mu_\theta,\sigma_\theta,\sigma_y,\mD,\mM_{1b})p(\mu_\theta,\sigma_\theta,\sigma_y|\mD,\mM_{1b}) \, \mathrm{d}\mu_\theta \mathrm{d}\sigma_\theta \mathrm{d}\sigma_y\\
&\approx \sum_{j=1}^{N_s} w_j p((\hat{x},\hat{y})|\mu_\theta^{(j)},\sigma_\theta^{(j)},\sigma_y^{(j)},\mD,\mM_{1b})
\end{aligned}
\end{equation}
where $p((\hat{x},\hat{y})|\mu_\theta^{(j)},\sigma_\theta^{(j)},\sigma_y^{(j)},\mD,\mM_{1b})$ can be evaluated analytically similar to the case in $\mM_{1a}$:
\begin{equation}
\begin{aligned}
&p((\hat{x},\hat{y})|\mu_\theta^{(j)},\sigma_\theta^{(j)},\sigma_y^{(j)},\mD,\mM_{1b})\\ 
=& \frac{1}{\sqrt{2\pi \left( (\sigma_y^{(j)}/\hat{x})^2 + (\tilde{\sigma}_\theta^{(j)})^2 \right)}}\mathrm{exp}\left( -\frac{(\tilde{\mu}_\theta^{(j)} - \hat{y}/\hat{x})^2}{2\left((\sigma_y^{(j)}/\hat{x})^2 + (\tilde{\sigma}_\theta^{(j)})^2\right)} \right)
\end{aligned}
\end{equation}

% Model 2a
\subsection{Zero noise HSFM, $\mM_{2a}$}
In this model, we assume the same Gaussian prior of $\theta_i$ for all $i = 1,\dots,N_D$, where $\theta_i$ corresponds to the data set $D_i \in \mD$. The hyperparameters $\vpsi = \{\mu_\theta,\sigma_\theta\}$ define the mean and the standard deviation of the Gaussian prior, respectively. By assuming a delta function for the likelihood, we can derive that for each data set $D_i$:
\begin{equation}
	\begin{aligned}
	p(\theta_i|D_i,\mu_\theta,\sigma_\theta,\mM_{2a}) &\propto p(D_i|\theta_i,\mM_{2a})p(\theta_i|\mu_\theta,\sigma_\theta,\mM_{2a})\\
	&= \prod_{j=1}^{N_{D_i}} \delta(y_{j,i} - \theta_i x_{j,i}) N(\theta_i|\mu_{\theta},\sigma_{\theta}^2)\\
	\text{i.e.,} \,\,\,\,\, p(\theta_i|D_i,\mu_{\theta},\sigma_{\theta},\mM_{2a}) &= 
	\begin{cases}
	1, \text{if there exists} \,\, \theta_i \,\, \text{such that} \,\, y_{j,i} = \theta_i x_{j,i} \,\, \text{for all} \,\, j = 1,\dots,N_{D_i}\\
	0, \text{otherwise}
	\end{cases}
	\end{aligned}
\end{equation}
And when the posterior equals 1, the evidence term $p(D_i|\mu_\theta,\sigma_\theta,\mM_{2a})$ calculated using Equation \ref{eq:A6} is:
\begin{equation}\label{eq:evid_M2a}
	p(D_i|\mu_\theta,\sigma_\theta,\mM_{2a}) = \frac{1}{\sqrt{2\pi}\sigma_\theta}\mathrm{exp}\left( -\frac{1}{2\sigma_\theta^2}\left(\frac{y_{1,i}}{x_{1,i}} - \mu_\theta \right)^2 \right) \cdot \prod_{j=1}^{N_{D_i}} \frac{1}{|x_{j,i}|}
\end{equation}
Otherwise, the evidence is zero. Note that the extra product of inverse of $x_{j,i}$ comes from the Jacobian of the delta likelihood function.

Similar to $\mM_{1b}$, we use $N_s$ samples from Monte Carlo Simulation to estimate the posterior of $\mu_\theta$ and $\sigma_\theta$:
\begin{equation}\label{eq:sigy_M2a}
	\begin{aligned}
	&p(\mu_\theta,\sigma_\theta|\mD,\mM_{2a}) \approx \sum_{j=1}^{N_s} w_j \delta(\mu_\theta - \mu_\theta^{(j)}) \delta(\sigma_\theta - \sigma_\theta^{(j)})\\
	\text{where} \,\,\,\,\, & (\mu_\theta^{(j)},\sigma_\theta^{(j)}) \sim p(\mu_\theta,\sigma_\theta|\mM_{2a}), w_j \propto \prod_{i=1}^{N_D}p(D_i|\mu_\theta^{(j)},\sigma_\theta^{(j)},\mM_{2a}), \sum_{j=1}^{N_s} w_j = 1
	\end{aligned}
\end{equation}
and the model evidence:
\begin{equation}
	p(\mD|\mM_{2a}) \approx \frac{1}{N_s} \sum_{j=1}^{N_s} \left(\prod_{i=1}^{N_D}p(D_i|\mu_\theta^{(j)},\sigma_\theta^{(j)},\mM_{2a})\right)
\end{equation}
In the HSFM, $\theta$ for future predictions (denoted as $\theta_0$ in Section \ref{sec:HSFM}), is independent of the data when $\vpsi$ is given. Hence, the marginalized posterior of $\theta$ depends only on the posterior of $\mu_\theta$ and $\sigma_\theta$ and the prior of $\theta$ given $\mu_\theta$ and $\sigma_\theta$:
\begin{equation}
	\begin{aligned}
	p(\theta|\mD,\mM_{2a}) &= \int \! p(\theta|\mu_\theta,\sigma_\theta,\mM_{2a})p(\mu_\theta,\sigma_\theta|\mD,\mM_{2a}) \, \mathrm{d} \mu_\theta \mathrm{d} \sigma_\theta\\
	&\approx \sum_{j=1}^{N_s} w_j N(\mu_{\theta}^{(j)},\sigma_{\theta}^{(j)})
	\end{aligned}
\end{equation}
As a result, the statistics of the marginalized posterior of $\theta$ can also be estimated based on the MCS samples:
\begin{equation}
	\begin{aligned}
	E[\theta|\mD] &= \int \! \theta p(\theta|\mD,\mM_{2a}) \, \mathrm{d} \theta \approx \sum_{j=1}^{N_s} w_j \mu_\theta^{(j)}\\
	Std[\theta|\mD] &\approx \sqrt{\left( \sum_{j=1}^{N_s} w_j \left( (\sigma_\theta^{(j)})^2 + (\mu_\theta^{(j)})^2 \right) \right) - E[\theta|\mD]^2}
	\end{aligned}
\end{equation}
To perform robust prediction of a new point $(\hat{x},\hat{y})$, we need to evaluate $p((\hat{x},\hat{y})|\mD,\mM_{2a})$. Again, we can use the posterior samples from Equation \ref{eq:sigy_M2a}:
\begin{equation}
	\begin{aligned}
	p((\hat{x},\hat{y})|\mD,\mM_{2a}) &= \int \! p((\hat{x},\hat{y})|\mu_\theta,\sigma_\theta,\mM_{2a})p(\mu_\theta,\sigma_\theta|\mD,\mM_{2a}) \, \mathrm{d}\mu_\theta \mathrm{d}\sigma_\theta\\
	&\approx \sum_{j=1}^{N_s} w_j p((\hat{x},\hat{y})|\mu_\theta^{(j)},\sigma_\theta^{(j)},\mM_{2a})
	\end{aligned}
\end{equation}
Note that an unique feature of the HSFM is that future predictions do not depend on the data once all hyperparameters are given. Hence, $p((\hat{x},\hat{y})|\mu_\theta^{(j)},\sigma_\theta^{(j)},\mM_{2a})$ does not have $\mD$ in the conditional part, and we can evaluate it using Equation \ref{eq:evid_M2a} by substituting $D_i$ with $(\hat{x},\hat{y})$.

% Model 2b
\subsection{Full HSFM, $\mM_{2b}$}\label{sec:A4}
In this model, we use the same setup as in $\mM_{2a}$ for the hyperparameters $\vec{\psi}$. By assuming a Gaussian distribution for the likelihood and the priors of $\theta_i$ for all $i = 1,\dots,N_D$ conditional on the hyperparameters, we can obtain analytical expressions similar to the one in $\mM_{1b}$. One difference is that in this model, we now work with multiple data sets $D_i \in \mD, i = 1,\dots,N_D$. Following a similar derivation as in $\mM_{1b}$, the posterior distribution $p(\theta_i|D_i,\mu_\theta,\sigma_\theta,\sigma_y,\mM_{2b})$ is Gaussian:
\begin{equation}\label{eq:postT2}
	\begin{aligned}
	&p(\theta_i|D_i,\mu_\theta,\sigma_\theta,\sigma_y,\mM_{2b}) = N(\theta_i|\tilde{\mu}_{\theta,i},(\tilde{\sigma}_{\theta,i})^2)\\
	\text{where} \,\,\, &\tilde{\sigma}_{\theta,i} = \frac{\sigma_{\theta}\sigma_y}{\sqrt{\vec{x}_i^T \vec{x}_i\sigma_{\theta}^2+\sigma_y^2}}, \, \tilde{\mu}_{\theta,i} = \frac{\vec{x}_i^T \vec{y}_i \sigma_{\theta}^2 + \mu_{\theta}\sigma_y^2}{\vec{x}_i^T \vec{x}_i\sigma_{\theta}^2+\sigma_y^2} = \left( \frac{\vec{x}_i^T \vec{y}_i}{\sigma_y^2}+\frac{\mu_{\theta}}{\sigma_{\theta}^2} \right)\tilde{\sigma}_{\theta,i}^2
	\end{aligned}
\end{equation}
and the evidence term $p(D_i|\mu_\theta,\sigma_\theta,\sigma_y,\mM_{2b})$ calculated using Equation \ref{eq:A6} is:
\begin{equation}\label{eq:evidT2}
	p(D_i|\mu_\theta,\sigma_\theta,\sigma_y,\mM_{2b}) = \frac{\tilde{\sigma}_{\theta,i}}{(2\pi)^{\frac{N_{D_i}}{2}} \sigma_y^{N_{D_i}}\sigma_{\theta}}\mathrm{exp}\left(-\frac{1}{2}\left( \frac{\vec{y}_i^T \vec{y}_i}{\sigma_y^2} + \frac{\mu_{\theta}^2}{\sigma_{\theta}^2} - \frac{\tilde{\mu}_{\theta,i}^2}{\tilde{\sigma}_{\theta,i}^2}\right) \right)
\end{equation}
Similar to $\mM_{1b}$, we use $N_s$ samples from Monte Carlo Simulation to estimate the posterior of $\mu_\theta$, $\sigma_\theta$ and $\sigma_y$:
\begin{equation}\label{eq:sigy_M2b}
	\begin{aligned}
	p(\mu_\theta,\sigma_\theta,\sigma_y|\mD,\mM_{2b}) &\propto p(\mD|\mu_\theta,\sigma_\theta,\sigma_y,\mM_{2b})p(\mu_\theta,\sigma_\theta,\sigma_y|\mM_{2b})\\
	&= \prod_{i=1}^{N_D}p(D_i|\mu_\theta,\sigma_\theta,\sigma_y,\mM_{2b})p(\mu_\theta,\sigma_\theta,\sigma_y|\mM_{2b})\\
	&\approx \sum_{j=1}^{N_s} w_j \delta(\mu_\theta - \mu_\theta^{(j)}) \delta(\sigma_\theta - \sigma_\theta^{(j)})\delta(\sigma_y - \sigma_y^{(j)})
	\end{aligned}
\end{equation}
\begin{equation*}
	\text{where} \,\,\,\,\, (\mu_\theta^{(j)},\sigma_\theta^{(j)},\sigma_y^{(j)}) \sim p(\mu_\theta,\sigma_\theta,\sigma_y|\mM_{2b}), w_j \propto \prod_{i=1}^{N_D}p(D_i|\mu_\theta^{(j)},\sigma_\theta^{(j)},\sigma_y^{(j)},\mM_{2b}), \sum_{j=1}^{N_s} w_j = 1
\end{equation*}
and the model evidence:
\begin{equation}
	p(\mD|\mM_{2b}) \approx \frac{1}{N_s} \sum_{j=1}^{N_s} \left(\prod_{i=1}^{N_D}p(D_i|\mu_\theta^{(j)},\sigma_\theta^{(j)},\sigma_y^{(j)},\mM_{2b})\right)
\end{equation}
Similar to $\mM_{2a}$, the marginalized posterior of $\theta$ can be estimated by:
\begin{equation}
	p(\theta|\mD,\mM_{2b}) \approx \sum_{j=1}^{N_s} w_j N(\theta|\mu_{\theta}^{(j)},(\sigma_{\theta}^{(j)})^2)
\end{equation}
As a result, the statistics of the marginalized posterior of $\theta$ and $\sigma_y$ can also be estimated based on the MCS samples:
\begin{equation}
	\begin{aligned}
	E[\theta|\mD] &= \int \! \theta p(\theta|\mD,\mM_{2b}) \, \mathrm{d} \theta \approx \sum_{j=1}^{N_s} w_j \mu_\theta^{(j)}\\
	Std[\theta|\mD] &\approx \sqrt{\left( \sum_{j=1}^{N_s} w_j \left( (\sigma_\theta^{(j)})^2 + (\mu_\theta^{(j)})^2 \right) \right) - E[\theta|\mD]^2}\\
	E[\sigma_y|\mD] &\approx \sum_{j=1}^{N_s} w_j \sigma_y^{(j)}\\
	Std[\sigma_y|\mD] &= \sqrt{\left( \sum_{j=1}^{N_s} w_j (\sigma_y^{(j)})^2 \right) - E[\sigma_y|\mD]^2}
	\end{aligned}
\end{equation}
To perform robust prediction of a new point $(\hat{x},\hat{y})$, we need to evaluate $p((\hat{x},\hat{y})|\mD,\mM_{2b})$. Again, we can use the posterior samples from Equation \ref{eq:sigy_M2b}:
\begin{equation}
	\begin{aligned}
	p((\hat{x},\hat{y})|\mD,\mM_{2b}) &= \int \! p((\hat{x},\hat{y})|\mu_\theta,\sigma_\theta,\sigma_y,\mM_{2b})p(\mu_\theta,\sigma_\theta,\sigma_y|\mD,\mM_{2b}) \, \mathrm{d}\mu_\theta \mathrm{d}\sigma_\theta \mathrm{d}\sigma_y\\
	&\approx \sum_{j=1}^{N_s} w_j p((\hat{x},\hat{y})|\mu_\theta^{(j)},\sigma_\theta^{(j)},\sigma_y^{(j)},\mM_{2b})
	\end{aligned}
\end{equation}
Similar to $\mM_{2a}$, we can evaluate $p((\hat{x},\hat{y})|\mu_\theta^{(j)},\sigma_\theta^{(j)},\sigma_y^{(j)},\mM_{2b})$ using Equation \ref{eq:evidT2} by substituting $D_i$ with $(\hat{x},\hat{y})$.

\end{appendices}

\bibliographystyle{plainnat}
\bibliography{HSM}

\end{document}